\documentstyle{article}
\newcommand{\newsec}{\setcounter{equation}{0}\section}%
\newcommand{\be}{\begin{equation}}
\newcommand{\ee}{\end{equation}}
\newcommand{\bea}{\begin{eqnarray}}
\newcommand{\eea}{\end{eqnarray}}
\newcommand{\beast}{\begin{eqnarray*}}
\newcommand{\east}{\end{eqnarray*}}

\def\cfel{\frac{c}{2}}
\def\pinf{+\infty}
\def\minf{-\infty}
\def\pminf{\pm\infty}
\def\Ran{\mbox{Ran$_g$}}
\def\hdir{H^{(D)}_{\o,L}}
\def\ids{{\cal N}}
\def\hid{\chi_{\D}}
\def\hie{\chi_{\{E\}}}
\def\hieo{\hie(H(\o))}
\def\pe{P(\{E\})}
\def\pdel{P(\D)}
\def\hilb{{\cal H}}
\def\subh{{\cal A}}
\def\mathrm#1{{\rm #1}}
\def\mathbf#1{{\bf #1}}
\def\xpp{{\mathrm{pp}}}
\def\xcont{{\mathrm{cont}}}
\def\xess{{\mathrm{ess}}}
\def\xac{{\mathrm{ac}}}
\def\xsc{{\mathrm{sc}}}
\def\xsing{{\mathrm{sing}}}
\def\Ppp{P_{\xpp}}
\def\Pcont{P_{\xcont}}
\def\Psing{P_{\xsing}}
\def\Pac{P_{\xac}}
\def\Psc{P_{\xsc}}
\def\ppp{{\p^{\xpp}}}
\def\pac{{\p^{\xac}}}
\def\psic{{\p^{\xsc}}}
\def\Mac{M_{\xac}}
\def\Msing{M_{\xsing}}
\def\Msc{M_{\xsc}}
\def\Mpp{M_{\xpp}}
\def\hpp{\hilb_{\xpp}}
\def\hcont{\hilb_{\xcont}}
\def\hac{\hilb_{\xac}}
\def\hsc{\hilb_{\xsc}}
\def\sess{\s_{\xess}}
\def\spp{\s_{\xpp}}
\def\scont{\s_{\xcont}}
\def\sac{\s_{\xac}}
\def\ssc{\s_{\xsc}}
\def\mpp{\m_{\xpp}}
\def\mcont{\m_{\xcont}}
\def\mac{\m_{\xac}}
\def\msing{\m_{\xsing}}
\def\msc{\m_{\xsc}}
\def\eltu{\ell^2(\mbox{$\Bbb Z$})}
\def\suminf{\sum_{n=-\infty}^{\infty}}
\def\nnu{\nonumber}
\def\goto{\rightarrow}
\def\goton{\goto n}
\def\gotoinf{\rightarrow\infty}
\def\Sch{Schr\"o\-din\-ger\ }
\def\para{\|}
\def\dd{\mbox{d}}
\def\a{\alpha}
\def\b{\beta}
\def\g{\gamma}
\def\d{\delta}
\def\e{\epsilon}
\def\ve{\varepsilon}

\def\l{\lambda}
\def\m{\mu}
\def\n{\nu}
\def\x{\xi}
\def\r{\rho}

\def\s{\sigma}
\def\t{\tau}
\def\f{\phi}
\def\vf{\varphi}

\def\p{\psi}
\def\o{\omega}

\def\D{\Delta}

\def\S{\Sigma}
\def\F{\Phi}
\def\P{\Psi}
\def\O{\Omega}
\def\tr{\mbox{tr}\,}
\def\start{\left(\begin{array}}
\def\fin{\end{array}\right)}
\def\opt{\left\{\begin{array}{ll}}
\def\finopt{\end{array}\right.}
\def\half{\frac{1}{2}}
\def\fourth{\frac{1}{4}}
\def\mup{\m_{\p}}
\def\mpe{\mup(E)}
\def\supp{\mbox{supp}\,}
\def\suppm{\supp\m}
\def\Zz{{\Bbb Z}}
\def\Rr{{\Bbb R}}
\def\Cc{{\Bbb C}}
\def\Qq{{\Bbb Q}}
\def\Nn{{\Bbb N}}
\def\xinb{x\in B:\ }
\def\mpx{(\dd\m/\dd m)(x)}
\def\einr{E\in\Rr:}
\def\Bbb{\mathbf}
\catcode`\@=11
\newcount\REFFcount \REFFcount=1
\def\refindent{\begin{itemize}}
\def\reff[#1]#2\eref{\let\\=\newline
     \def\(##1){{``\sl ##1''}}\def\<##1>{{\sl ##1}}%
     \def\>##1<{{\bf ##1}}\def\[##1]{{\ninepoint\sl [\ ##1\ ]}}%
     \item[{[#1]}]\ignorespaces #2\par}
     \let\eref\relax
\def\beginref#1{\subsection*{\romannumeral\REFFcount)\enspace #1}
    \global\advance\REFFcount by 1\begingroup\refindent}
\def\endref{\end{itemize}\endgroup}

\catcode`\@=12

\begin{document}
\title{{\flushleft{\small {\rm Appeared in: {\em Beyond
Quasicrystals}, eds. F. Axel and D. Gratias, Springer-Verlag \& Les Editions de Physique (1995) pp. 481-549.}\\}}\vspace{2cm}
\Sch difference equation with deterministic ergodic potentials}
\author{Andr\'as S\"ut\H{o}\thanks{Present and permanent address:
        Research Institute for Solid State Physics, POB 49,
        H-1525 Budapest 114, Hungary.
        Email: suto@szfki.hu}\\
        Institut de Physique Th\'eorique\\
        Ecole Polytechnique F\'ed\'erale de Lausanne\\
        CH-1015 Lausanne, Switzerland}
\date{}
\maketitle
\begin{abstract}
We review the recent developments in the theory of the one-dimensional
tight-binding \Sch equation for a class of deterministic ergodic potentials.
In the typical examples the potentials are
generated by substitutional sequences,
like the Fibonacci or the Thue-Morse sequence. We concentrate on rigorous
results which will be explained rather than proved. The necessary mathematical
background is provided in the text.
\end{abstract}

\load{\tiny}{\rm}
\section*{Contents}
\begin{enumerate}
\item Introduction
\item Main examples
\item General results on the \Sch difference equation
\begin{itemize}
\item[3.1] Basic observations
\item[3.2] Transfer matrices
\item[3.3] Lyapunov exponent
\item[3.4] Scattering problem: Landauer resistance
\end{itemize}
\item \Sch equation with periodic potentials
\item Spectral theory
\begin{itemize}
\item[5.1] \Sch operator, $\eltu$-space and spectrum
\item[5.2] Point spectrum
\item[5.3] Cantor sets
\item[5.4] Continuous spectrum
\item[5.5] Spectral projection
\item[5.6] Measures
\item[5.7] Cantor function
\item[5.8] Spectral measures and spectral types
\item[5.9] A spectral measure for $H_0$
\item[5.10] $\eltu$ versus $\ell^2(\Nn)$
\item[5.11] Asymptotic behaviour of generalized eigenfunctions. Subordinacy
\end{itemize}
\item \Sch equation with strictly ergodic potentials
\begin{itemize}
\item[6.1] Strict ergodicity
\item[6.2] The spectrum of $H(\o)=H_0+V(\o)$
\item[6.3] Integrated density of states
\item[6.4] IDS and Lyapunov exponent
\item[6.5] Results on the set $\{E: \g(E)=0\}$
\item[6.6] The role of periodic approximations
\item[6.7] Gordon-type theorems
\item[6.8] Kotani theorem for potentials of finite range
\item[6.9] Gap labelling
\end{itemize}
\item \Sch equation with Sturmian and substitutional potentials
\begin{itemize}
\item[7.1] Fibonacci potential
\item[7.2] General Sturmian potentials
\item[7.3] Period doubling potential
\item[7.4] Thue-Morse potential
\item[7.5] Systematic study of substitutional potentials
\end{itemize}
\item Solutions to the problems
\end{enumerate}
\begin{itemize}
\item[] References
\end{itemize}
\newpage
\section{Introduction}
The one-dimensional \Sch equation
\[(-d^2/dx^2+V(x))\psi(x)=E\psi(x)\]
has long been serving as a laboratory for
studying a large variety of problems in Quantum Mechanics. Most often it appears
as the eigenvalue equation of a one-electron energy operator, and then $V(x)$
means the potential energy of the electron. In Atomic or Molecular Physics
$V(x)$ is typically fast-decaying or vanishes outside a bounded set, while in
applications to Solid State Physics it is an extended function. Our concern
will be the \Sch equation of Solid State Physics in the tight-binding
approximation: the
real axis is replaced by $\Bbb Z$, the lattice of integers, and the
differential operator by the second order difference operator.
The tight-binding \Sch equation describes an electron in a lattice
in which the attraction of the ion cores slows down the interatomic
motion of the electron:
the kinetic energy has an {\em a priori} upper bound which does not exist in
the continuous case.

The potentials we are going to consider are on the midway between periodic and
random; in physics literature they are sometimes called `aperiodic' or
`disordered deterministic'. They all fall into the family of the so-called
strictly ergodic sequences. A typical subfamily which appeared regularly
in the lectures of this Winter School are the substitutional sequences as, for
example, the Fibonacci sequence. As we will see later, the periodic and the
uniformly almost periodic sequences are also strictly ergodic, but a random
potential is usually `only' ergodic.

In this field physicists and mathematicians have been working parallelly
during about the last fifteen years. The ambition of this course is to review
and explain the contribution of the latter for an audience unfamiliar with
the underlying mathematics.
Most of the time, mathematicians consider the
\Sch equation directly on the
infinite chain $\Bbb Z$ or on the semi-infinite chain $\Bbb N$. The equation is
then the `eigenvalue equation' of an infinite matrix which
is named after Jacobi. If the diagonal (the potential) is an ergodic sequence,
the spectrum happens not to be a {\em discrete} point spectrum.
It is rather often - at least within the family of
potentials we are going to consider - singularly continuous and is, as an
ensemble of points, a (generalized) Cantor set. These notions and many others,
necessary for the understanding of the specific results will be introduced in
the forthcoming sections.

We start, in Section 2, by the presentation of the main examples
and the definition of strict ergodicity. The subject is developed in Sections
3, 4, 6 and 7.
In Section 3 we give the
basic ingredients of the theory of discrete \Sch equation in one dimension
and introduce the notions of the transfer matrix and Lyapunov exponent.
A subsection is devoted to the scattering problem and Landauer resistance.
Section 4 contains some of the fundamental results for periodic potentials.
Section 6 summarizes the general results
obtained for one-dimensional \Sch operators with strictly ergodic potentials.
Section 7 is a review of our actual knowledge about models with specific
potentials, given by Sturmian and substitutional sequences.

Section 5 is an introduction into spectral theory. It is far more the longest
part of the course, with the definitions of
spectrum, essential spectrum, eigenvalue, point and continuous spectrum,
spectral projections, measures and spectral measures, Cantor sets etc.,
illustrated with examples.
Subordinacy and the classification of the
spectrum according to the asymptotic behavior of the solutions of the \Sch
equation is also discussed here.

The widely studied
Almost-Mathieu or Harper equation is not presented in full detail
but appears in examples throughout the whole text, and papers on it are included
among the references.

Some more or less simple
facts are announced  in the form of problems. The reader may
only retain the statements or, if he wishes, check his understanding by proving
them. The solutions can be found in Section 8.

These notes are not written in a purely mathematical style. Proofs, unless
they are really elementary, are replaced by explanations. This suffices for the
physicist who wishes to understand the results but has no ambition to
provide mathematical proofs of his own. The interested reader can find excellent
monographs which present the subject with full mathematical rigor. Also, one
should remember that nothing can replace the lecture of the original research
papers. A rather detailed, although certainly incomplete, list of references
is provided at the end of the notes.
I thank Fran\c{c}ois Monti for helping me to construct this list.

It is a pleasure to thank Fran\c{c}ois Delyon, Fran\c{c}ois Monti and Charles
Pfister, who accepted to read the manuscript; their remarks brought considerable
improvement to the final text.
\newsec{Main examples}
The tight-binding \Sch equation in one dimension is written as
\be\label{Sch}
\p_{n-1}+\p_{n+1}+V_n\p_n=E\p_n
\ee
Here $n$ runs over the integers, $V=\{V_n\}$ is a real sequence (the potential),
$E$ is a real number (the energy), although sometimes it is useful to take it
complex. The problem is to find the `physically allowed' values of $E$, that is,
those admitting at least one `not too fast increasing' solution, and to describe
these solutions. The following potentials will be considered:

(1) $V_n=V(\o)_n=\l\cos2\pi(n\a+\o)$, $\l\neq0$. The corresponding discrete
\Sch equation is called the Almost-Mathieu
equation (because the continuous version is the Mathieu equation). The case
$\l=2$ is the Harper equation.

(2) $V_n=V(\o)_n=\l X_A(n\a+\o)$, where $\l\neq0$, $A$ is the union of
semi-open intervals of $[0,1)$
and $X_A(x)=1$ if the fractional
part of $x$ falls into $A$ and is zero otherwise. These are the so-called
circle-potentials. An important special case is when $A$ is a semi-open
interval of length $\a$: The $\{0,1\}$-valued sequences
\bea\label{Sturm}
X_{[0,\a)}(n\a+\o)=\lfloor n\a+\o\rfloor-\lfloor(n-1)\a+\o\rfloor\nonumber\\
X_{(0,\a]}(n\a+\o)=\lceil n\a+\o\rceil-\lceil(n-1)\a+\o\rceil
\eea
for irrational $\a$
are called Sturmian sequences. The notations $\lfloor\cdot\rfloor$ and
$\lceil\cdot\rceil$ mean rounding downwards and upwards, respectively.
If $\a=(\sqrt{5}-1)/2$, we obtain the Fibonacci sequence. Generalized Fibonacci
sequences (obtained through the substitution $\x(a)=a^nb$ and $\x(b)=a$ with
$n\geq1$) are also Sturmian. According to an
equivalent definition, Sturmian sequences are the sequences with minimal
complexity among the non-ultimately-periodic sequences (see e.g. [1],
[23]).

(3) $V$ is a substitutional sequence:\\
Let $\x$ be a primitive substitution ([114]) on a finite alphabet
${\cal A}=\{a_1,...,a_r\}$,
$f$ a non-constant real function on $\cal A$. Suppose that $\x(a)=...a$
and $\x(b)=b...$ for some $a, b\in \cal A$ and let $\o=uv$ where $u$ and $v$ are
the left- and right-sided fixed points of $\x$:
$u=\x^{\infty}(a)=...a$ and $v=\x^{\infty}(b)=b...$.
Then $V(\o)_n=f(\o_n)$.
A comment on the concatenation $uv$ will follow in Section 6.
Generalized Fibonacci
sequences have a natural definition on $\Zz$ via the formula for Sturmian
sequences.

The potentials defined in (1)-(3) are {\em strictly ergodic}, that is,
{\em minimal} and {\em uniquely ergodic}:

(i) $V$ is minimal.\\
Let $T$ be the left shift, and denote $V(T\o)$ the shifted potential:
$V(T\o)_n=V(\o)_{n+1}$. The notation ($T$ applied to $\o$) may seem curious,
however, in all cases $T\o$ can be given a well-defined meaning.
For potentials (1) and (2)
\[T\o=\o+\a\;\mbox{(mod $1$)}\]
In case (3), $T\o$ is the left-shifted sequence,
\[(T\o)_n=\o_{n+1}\]
The orbit of $V(\o)$, i.e., the sequence of translates, $V(T^k\o)$ for
$k=...,-1,0,1,...$, has
pointwise convergent subsequences:
$W=\{W_n\}$ is called a pointwise limit of translates of $V(\o)$ if
for a suitably chosen sequence
$\{k_i\}$, $V(T^{k_i}\o)_n$ tends to $W_n$ for every $n$. Clearly,
$W=V(\o')$, where $\o'$ is the limit of $T^{k_i}\o$ in cases (1) and (2) and
is a pointwise limit of translates of $\o$ in case (3).
%

The orbit of $V$ together with all its pointwise limits
is called the {\em hull} of $V$ and is denoted by $\O(V)$.
This is a kind of closure of the set of translates. The hull can be
constructed for any infinite sequence. If $V$ is periodic with period length
$L$, the hull is a set of $L$ elements.
The potentials of (1) and (2) are periodic if
$\a$ is rational. If $\a$ is irrational, the hull of $V=V(\o=0)$ is
$\{V(\o)\}_{\o\in[0,1)}$! This is the typical situation: the hull is much larger
(uncountable) than the set of translates (countable). If $S$ is an arbitrary
sequence and $S'$ is a pointwise limit of translates of $S$, the hull of $S'$
is usually different from the hull of $S$.  (Example: $S_n=1-1/n, n=1,2,...$.
The translates have a single pointwise limit, $S'_n\equiv 1$. The only element
of $\O(S')$ is $S'$ itself.)

The potentials defined in (1)-(3) have the property
that for any $W\in\O(V)$, $\O(W)=\O(V)$. This is what the minimality of $V$
means. The $V$-independent hull is denoted by
$\O$. Clearly, every element of $\O$ is minimal and therefore the `flow'
$(\O,T)$ itself can be called minimal: every trajectory
is dense in $\O$ (its closure is $\O$) or equivalently, the only closed
and shift-invariant subsets of $\O$ are the empty set and $\O$.

{\em Problem 1.} (Cf. Queff\'elec [113]) Let $s=\{s_n\}_{-\infty}^\infty$
be a bounded complex sequence, $\O(s)$ its hull,
\[\O(s)=\overline{\{T^ks\}_{k=-\infty}^{\quad\infty}}\]
where the bar means closure with respect to pointwise convergence.
The following are equivalent.\\
1. $s$ is minimal, i.e., for every $t\in\O(s)$, $\O(t)=\O(s)$.\\
2. For every $t\in\O(s)$, $s\in\O(t)$.\\
3. $s$ is almost-periodic in the metric
\[d(s,t)=\sum_{n=-\infty}^\infty 2^{-|n|}|s_n-t_n|\;\]
i.e., for each $\ve>0$ there exists an $\ell_\ve<\infty$ and a sequence
$\{n_k\}_{k=-\infty}^{\quad\infty}$ of integers with
gaps bounded by $\ell_\ve$,
$0<n_{k+1}-n_k<\ell_\ve$, such that for all $k$
\[d(T^{n_k}s,s)=\sum_{n=-\infty}^\infty 2^{-|n|}|s_{n+n_k}-s_n|\leq\ve\;.\]

A sequence is called {\em recurrent} if it is the pointwise limit of its
own translates. A minimal sequence is recurrent but
a recurrent sequence may not be minimal.
A counterexample is provided by the sequence of digits $1234567891011121314...$,
obtained by concatenating the positive integers, written, e.g., in base $10$.
It is the pointwise limit of its own left shifts, but is not minimal: words
are repeated with increasing gaps. The hull of this sequence is the set of all
sequences composed of the digits $0,1,...,9$.

The different meanings, in different cases, of the label $\o$ should not
disturb the reader. In
(1) and (2), $\o$ is a number of the interval $[0,1)$;
in (3) it is an element of the hull of $uv$, and
$uv$ is also a minimal sequence.

\vspace{2mm}

(ii) $V$ is uniquely ergodic.\\
Recall that the dynamical system
$(\O,T,\r)$, where $\r$ is a $T$-invariant probability measure,
is ergodic if for any $\r$-integrable function $f$
\[\lim_{N\gotoinf}N^{-1}\sum_{n=1}^N f(T^n\omega_0)=\int_\O f(\o)\dd\r(\o)\]
for $\r$-almost every $\o_0$ (i.e., apart from a set of $\o_0$'s of
zero $\r$-measure). Choosing $f$ to be the
characteristic function of a set $A$, it is seen
that ergodicity implies that (in fact, is equivalent to) $\r(A)=0$ or $1$
for every $T$-invariant set $A$.

The unique ergodicity of $V$ means that there exists a unique
shift-invariant probability measure $\r$ on $\O(V)$ ($\r(T^{-1}A)=\r(A)$
for any measurable $A\subset\O$).

{\em Problem 2.} If $s$ is a uniquely ergodic sequence with shift-invariant
probability measure $\r$ then $(\O(s),T,\r)$ is ergodic.

In case (1) (Almost-Mathieu), the most natural way to obtain this measure
is to define, for $0\leq a<b<1$,
\be\label{prob}
\r(A_{ab})=b-a\ \ ,\ \ \mbox{where $A_{ab}=\{V(\o)\}_{a<\o<b}$}
\ee
and to extend $\r$ to countable unions and intersections of the sets $A_{ab}$
by using the properties of measures.

For substitutional potentials
the usual construction is to assign probabilities to the
so-called {\em cylinder sets} and to extend $\r$ to countable unions
and intersections of cylinders.
If $v=(v_1,v_2,...,v_k)$ is a finite collection of
real numbers and n is an integer,
the cylinder $[v,n]$ is a family of potentials taken from the hull,
\be\label{cyl} [v,n]=\{V\in\O: V_{n+i}=v_i, i=1,\ldots,k\}\ee
Noticing that the relative frequency of the `word'
$v$ is the same in all $V\in\O$, the probability $\r([v,n])$ can be defined as
this relative frequency.

For circle potentials a shift-invariant $\r$ can be derived naturally either
through Eq. (\ref{prob}) or through word frequencies. For the Almost-Mathieu
potential cylinders can also be defined (by replacing, in Eq. (\ref{cyl}),
$V_{n+i}=v_i$ by $V_{n+i}\in I_i$, where $I_i$ is an interval),
and one can obtain a $T$-invariant
probability measure through them. Uniqueness implies that
the different constructions yield the same measurable sets and measures.


Physicists never worry (do not even have to know) about the hull;
mathematicians all the time struggle
with it. The reason is that in laboratory or computer experiments one
deals with finite samples, and these
may come from any element of the hull: therefore it is sufficient to keep just
one. (However, the physicist's sample averaging is nothing else than
averaging over the hull with the measure $\r$.)
The mathematician is naturally led to the notions of hull and strict
ergodicity by working on the infinite system. He (she) has to consider $V$ as a
$\r$-distributed random variable on $\O$. The results she (he) obtains are valid
either for a given $V\in\O$ or for every $V\in\O$ or, most often, for
$\r$-a.e. (almost every) $V\in\O$, i.e.,
$\r(\{V\in\O$: the result is not valid$\})=0$. Notice that the distinction
between `every' and `almost every' is by no means harmless. Indeed, because
$\r$ is a continuous probability measure (that is, $\r(\{V\})=0$ for every
one-point subset of $\O$), any countably infinite subset of the hull has
zero probability. A surprising new result about an even larger (uncountable)
set of vanishing $\r$-measure, for which a claim (localization) is not valid,
will be quoted in Section 5.2.

{\em Problem 3.} If $\r$ is a shift-invariant probability measure on $\O$
and the trajectory of every point is infinite then $\r$ is continuous.
\newsec{General results on the \Sch difference equation}
\subsection{Basic observations}
The first two terms of Eq.(\ref{Sch}) correspond to the action of the
kinetic energy operator on the wave function, and
any physicist would prefer to replace them by
\[-\p_{n-1}-\p_{n+1}+2\p_n\]
This is, however, a difference of no importance: Omitting $2\p_n$ means shifting
the zero of the energy by 2, and changing the sign of the first two entries
in Eq.(\ref{Sch}) amounts to a unitary transformation, thus leaving the
spectrum of the energy operator unchanged.
Therefore a global change of sign of $V$ is also of no
importance: we get the mirror image of the spectrum. This simple fact
can be expressed in a more shocking form: If $V$ gives rise to a
localized state at energy $E$, $-V$ will give rise to a localized state at
energy $-E$. Or still in another form: There can exist extended states at
energies everywhere below the potential, $E<V_n$ for all $n$.

Without saying otherwise, we always consider the \Sch equation on the
infinite lattice $\Bbb Z$.
Equation (\ref{Sch}) is a homogeneous linear second order difference equation.
As a consequence, for any (complex) $E$ the solutions form a two-dimensional
linear vector space:
\begin{itemize}
\item If $\p^1$ and $\p^2$ solve (\ref{Sch}), then $\p=c_1\p^1+c_2\p^2$ solves
it for any complex $c_1$ and $c_2$.
\item There {\em exist} two linearly independent solutions $\p^1$ and $\p^2$.
\end{itemize}
Two nontrivial ($\neq0$) solutions are linearly {\em dependent\/} if and only
if they are constant multiples of each other. When saying that Eq.(\ref{Sch})
has a unique solution of some property, we will understand `unique linearly
independent'.

Writing (\ref{Sch}) for $\p^1$ and $\p^2$ (not necessarily linearly
independent), multiplying the first equation by $\p^2_n$ and the second one by
$\p^1_n$ and subtracting we find that
\[
\p^1_{n+1}\p^2_n-\p^1_n\p^2_{n+1}=\p^1_n\p^2_{n-1}-\p^1_{n-1}\p^2_n\\
=\cdots=\p^1_1\p^2_0-\p^1_0\p^2_1
\]
This constant is the Wronskian and will be denoted by $W[\p^1,\p^2]$.
As a consequence,
\begin{itemize}
\item $\p^1$ and $\p^2$ are linearly independent solutions if and only if
      $W[\p^1,\p^2]\neq0$.
\item There can be {\em at most} one solution going to zero
as $n\gotoinf$. If $\p^1$ is such a solution, then for any other (linearly
independent) solution $\p$
\[|\p_n|^2+|\p_{n+1}|^2\gotoinf \mbox{  as  } n\gotoinf\]
The same conclusion holds for the limit $n\goto -\infty$. As a consequence,
there can be {\em at most} one solution decaying on both sides, and if such a
solution exists, all the others blow up on both sides.
\end{itemize}
\subsection{Transfer matrices}
We may rewrite the \Sch equation in the redundant vectorial form
\beast
\P_n=\left(\begin{array}{cc}E-V_n & -1\\1 & 0\end{array}\right)\P_{n-1}
\equiv T_n\P_{n-1}
\east
where
\be\label{vector}
\P_n=\left(\begin{array}{c}\p_{n+1}\\ \p_n\end{array}\right)
\ee
By iterating,
\[\P_n=T_nT_{n-1}\cdots T_1\P_0\equiv T_{1\goton}\P_0\]
and similarly,
\[\P_{-n}=T_{-n+1}^{-1}T_{-n+2}^{-1}\cdots T_0^{-1}\P_0=T_{0\goto-n+1}\P_0\]
The matrices $T$ are called transfer matrices. Their determinant is $1$:
this is clear for $T_n$ and its inverse and holds for the others because
the determinant of a product of matrices is the product of the determinants.

The determinant of $T_{1\goton}$ can be seen as the Wronskian of the two
`standard' solutions $\p^1$ and $\p^2$ with initial conditions
\beast
\begin{array}{rl}\p^1_1=1 & 0=\p^2_1\\ \p^1_0=0 & 1=\p^2_0\end{array}
\east
In terms of these solutions the transfer matrix can be written as
\bea\label{trans}
T_{1\goton}=T_{1\goton}\left(\begin{array}{cc}
\p^1_1 & \p^2_1\\ \p^1_0 & \p^2_0\end{array}\right)=
\left(\begin{array}{cc}
\p^1_{n+1} & \p^2_{n+1}\\ \p^1_n & \p^2_n\end{array}\right)
\eea
showing that indeed,
\[\det T_{1\goton}=W[\p^1,\p^2]=1\]

The general form of the transfer matrix from site $m$ to site $n$ is
\be\label{abcd}
T_{m\goton}=A=\start{cc}a&b\\c&d\fin\ \ ,\ \ ad-bc=1
\ee
Here $a,b,c,d$ are polynomials of $E$ with real coefficients and of
degree $\leq |n-m|+1$. In particular,
\[ \tr A=a+d=E^{|n-m|+1}+\cdots\]

For real energies the transfer matrices are elements of the group
SL(2,$\Bbb R$). The characteristic equation of a matrix $A$ in this group reads
\[\det(\l-A)=\l^2-\l\,\tr A+1=0\]
Therefore the two roots satisfy the relations $\l_1\l_2=1$ and
$\l_1+\l_2=\tr A$. We can distinguish four cases:
\begin{enumerate}
\item The two roots are real, $|\l_1|=|\l_2^{-1}|>1$, the corresponding
      eigenvectors are also real and nonorthogonal, in general.
      $|\tr A|>2$, the matrix is {\em hyperbolic}.\\
      For example, the `hyperbolic rotations'
      \[\start{cc}\cosh x & \sinh x\\ \sinh x & \cosh x\fin\]
      are special (symmetric) hyperbolic matrices. $\l_{1,2}=\exp(\pm x)$, the
      corresponding eigenvectors are the transposed of $(1\ \ \pm 1)$.
\item $\l_1=\l_2^*=e^{i\vf}$, $\vf\neq n\pi$. There are two complex
      eigenvectors. $\tr A=2\cos\vf$, the matrix is {\em elliptic}.
      As an example, the ordinary rotation matrix
      \[\start{cc}\cos x & -\sin x\\ \sin x & \cos x\fin\]
      is elliptic. $\l_{1,2}=\exp(\pm ix$), the eigenvectors are the transposed
      of $(1\ \ \mp i)$.
\item $\l_1=\l_2=\pm 1$. There is a unique, real, eigenvector (the geometric
      multiplicity is $1$). $|\tr A|=2$, the matrix is {\em parabolic}.
      An example is
      \[\left(\begin{array}{cc}1 & b\\0 & 1\end{array}\right)\]
      with $b\neq 0$. The unique eigenvector is the transposed of $(1\ \ 0)$.
\item $A=\pm I$ where $I$ is the unit matrix.
\end{enumerate}

Some assertions will concern the norm of transfer matrices. The natural matrix
norm
\[ \para A\para=\sup\para Au\para/\para u\para\]
associated with the vector norm $\para u\para=(|u_1|^2+|u_2|^2)^{\half}$
satisfies
\bea\label{norm}
\|A\|^2=\mbox{max. eigenvalue of}\ A^*A
=\half\para A\para_2^2+\left(\fourth\para A\para_2^4-1\right)^{\half}
=\para A^{-1}\para^2\geq 1\nnu\\
\eea
Here $\para A\para_2$ is
the trace norm,
\be\label{norm2}
\para A\para_2^2=\tr A^*A=a^2+b^2+c^2+d^2
=\para A^{-1}\para_2^2\geq2
\ee
(cf. Eq.\ref{abcd}), which proves also
the last inequality in (\ref{norm}).

Let us return to the trace of the transfer matrix as a function of $E$.
\[\tr T_{1\goton}(E)=E^n+\cdots=p_n(E)\]
is an $n$th degree polynomial with $n$ {\em distinct real\/} roots and the
property that
\[\frac{\mbox{d}}{\mbox{d}E}p_n(E)\neq 0\ \ \mbox{if}\ \ |p_n(E)|<2\]
In Figure 1a we represented the generic situation for $n=5$: There are $n$
disjoint intervals of the energy within which the transfer matrix is elliptic,
separated by intervals of hyperbolicity. Figure 1b shows an atypical case
for $n=4$: Some intervals of hyperbolicity disappear because the derivative
of the trace vanishes at the same $E$ where the trace
takes on the value $2$ or $-2$.

{\em Problem 1.} For $A\in$SL(2,$\Rr$), show that $|\tr A^k|\leq2$ if and only if
$|\tr A|\leq2$. Plot $\tr T_{1\goton}^3(E)$ on the top of Figure 1.
\subsection{Lyapunov exponent}
Consider
\[\g_n=\frac{1}{|n|}\ln\|T_{1\goton}\|\]
This number is nonnegative (because the norm is $\geq 1$) and depends on $E$
and the parameters of the potential. From $\g_n$ we obtain,
as $n\goto\pm\infty$,
\[\overline{\g}_{\pm}=\limsup_{n\goto\pm\infty}\g_n\ \ ,\ \
\underline{\g}\,_{\pm}=\liminf_{n\goto\pm\infty}\g_n\]
We find the same numbers by using the trace- or any other norm.
In fortunate cases the limits exist (i.e., $\limsup$ and $\liminf$ coincide)
and give $\g_{\pm}=\g_{\pm}(E,V)$. These we call the Lyapunov exponents. If
$\g_+=\g_-$, the common value is denoted by $\g(E,V)$.

To elucidate the meaning of the Lyapunov exponent, recall that
\[\para T_{1\goton}\para_2^2=\para\P^1_n\para^2+\para\P^2_n\para^2\]
cf. equations (\ref{vector}), (\ref{trans}) and (\ref{norm2}).
Let, e.g., $n\goto+\infty$
and suppose that $\g_+$ exists. If $\g_+=0$, every solution
is subexponential. Suppose that $\g_+>0$. Then asymptotically
\[\ln\para\P^i_n\para\asymp n\g_+\]
for $i=1$ or $2$ or both and,
since any solution is a linear combination of the two standard ones,
the same asymptotics is valid for all solutions, except perhaps a unique one
(which is asymptotically smaller).
The Multiplicative Ergodic Theorem
due to Osceledec [104] and Ruelle [123] asserts that there is indeed
a unique solution $\p$ with an exponential decay:
\[\ln\para\P_n\para\asymp -n\g_+\]

If both of $\g_{\pm}$ are positive, it may happen that the unique solution
decaying exponentially to the left coincides with the unique solution decaying
exponentially to the right: we have an exponentially localized state.
If this is not the case, for the given energy all solutions increase
exponentially at least in one direction. Such an energy is considered as
`physically forbidden' and the solution physically irrelevant,
because there can be no electronic state inside the
solid, corresponding to it.

Let us make this statement somewhat more explicit.
We may ask: Provided that the electron is in the
interval $[-L,L]$, what is the probability of finding it at site $n$? The
answer is $|\p_n|^2/\sum_{k=-L}^L|\p_k|^2$. This number is of the order of
unity close to $L$ or $-L$ and exponentially decaying towards the inside.
So in any finite sample, at the given energy, the electron cannot penetrate
into the bulk, it is exponentially localized at the boundaries.
Also, one can learn from
generalized eigenfunction expansion that solutions increasing
faster than any power of $|n|$ will not appear in the expansion
of wave packets or computation of matrix elements
and are therefore of no relevance to Physics.
\subsection{Scattering problem: Landauer resistance}
Suppose we have a finite sample: a potential given at $n=1,...,L$. Our aim is
to discuss scattering of an electron of energy $E$ on the sample. We devise
our scattering experiment so as to have access to all energies.
To this end, the potential is extended to $\Bbb Z$ with constant values,
\[V_n=\opt\o_1 & \ \ \mbox{if}\ \ n\leq 0\\ \o_2&\ \ \mbox{if}\ \ n>L\finopt\]
chosen so that $|\o_1-\o_2|<4$. Then the scattering problem is defined for
energies satisfying the inequalities
\[|E-\o_1|<2\ \ ,\ \ |E-\o_2|<2\]
(Notice that $E<\o_i$ is allowed, see the remark in the first paragraph of
Section 3.1.) The general solution has the form
\be\label{scat}
\p_n=\opt Ae^{ik_1n}+Be^{-ik_1n},&\ \ n\leq 1\\
          Ce^{ik_2n}+De^{-ik_2n},&\ \ n\geq L\finopt
\ee
(the plane waves extend to $n=1$ and $n=L$) where the real wave numbers $k_i$
are related to $E$ and $\o_i$ via the equations
\[ 2\cos k_i=E-\o_i\ \ ,\ \ i=1,2\]
There is no loss of generality, if we suppose that $0<k_i<\pi$.
With this convention,
the scattering problem amounts to set $A=1$, $B=r$, $C=t$ and $D=0$ and
to solve for $t$ and $r$.
The Landauer resistance of the sample of length $L$ is defined as
$R_L=|r|^2/|t|^2$.

$\p$ with the scattering boundary condition is a complex solution of the
\Sch equation. Therefore, its complex conjugate is a linearly independent
solution to the same (real) energy.
The Wronskian between $\p$ and $\p^*$ is a constant times the
current density: its constancy physically means current conservation.
Compute the Wronskian at $n=0$ and at $n=L$. One finds
\[W[\p,\p^*]=2i(1-|r|^2)\sin k_1=2i|t|^2\sin k_2\]
so that
\[ |t|^2\sin k_2=(1-|r|^2)\sin k_1\]
This relation is well-known in the special case of $k_1=k_2$.
Because the sines are positive, $|r|$ cannot exceed $1$.

We want to express $R_L$ in terms of the elements $a,b,c,d$ of the transfer
matrix $T_{1\goto L}$. (Here we use the notations of Eq.(\ref{abcd})).
This can be done directly by solving the linear problem
\[\start{c}te^{ik_2(L+1)}\\te^{ik_2L}\fin=
\start{cc}a&b\\c&d\fin\start{c}e^{ik_1}+re^{-ik_1}\\1+r\fin \]
but there is a more elegant way to do it.
We can express the Landauer resistance also in terms of
another type of transfer matrix, usually applied for the \Sch equation in the
continuum. This transfer matrix,
\[Y=\start{cc}\a&\b\\ \g&\d\fin\]
connects the coefficients
$A$ and $B$ to $C$ and $D$, see Eq.(\ref{scat}):
\[\start{c}C\\D\fin=Y\start{c}A\\B\fin\]
In contrast with $T_{1\goto L}$, $Y$ depends on $k_1$ and $k_2$.
Let now $\p$ denote the solution with $A=1$ and $B=0$; it follows that $C=\a$
and $D=\g$. $\p^*$, the complex conjugate of $\p$, is also a solution with
$A=0$ and $B=1$ and, thus, with $C=\b$ and $D=\d$. We find therefore
\[\g=\b^*\ \ ,\ \ \d=\a^*\]
Furthermore, the Wronskian of $\p$ and $\p^*$ computed at $n=0$ and at $n=L$
yields
\[W[\p,\p^*]=2i\sin k_1=2i(|\a|^2-|\b|^2)\sin k_2\]
and thus
\[\det Y=|\a|^2-|\b|^2=\sin k_1/\sin k_2\]
The scattering problem
\[\start{c}t\\0\fin=Y\start{c}1\\r\fin\]
trivially gives
\be\label{rt}
r=-\b^*/\a^*\ \ ,\ \ t=\frac{1}{\a^*}\frac{\sin k_1}{\sin k_2}
\ee
from which
\[R_L=|\b|^2\frac{\sin^2k_2}{\sin^2k_1}\]
Another expression of $R_L$ in terms of the trace norm of $Y$ can also be
obtained.
\[\para Y\para^2_2=2|\a|^2+2|\b|^2=4|\b|^2+2\frac{\sin k_1}{\sin k_2}\]
With this we find
\[R_L=\fourth\para Y\para_2^2\frac{\sin^2k_2}{\sin^2k_1}-\half\frac{\sin k_2}
{\sin k_1}\]
These formulas are frequently used with $k_1=k_2$, see e.g. [46], [85],
[82].
It remains to connect the two different kinds of transfer matrices. This can
be done by simple inspection and results
\[\start{cc}e^{ik_2(L+1)}&e^{-ik_2(L+1)}\\e^{ik_2L}&e^{-ik_2L}\fin Y=
T_{1\goto L}\start{cc}e^{ik_1}&e^{-ik_1}\\1&1\fin \]
Solving this equation for $\a$ and $\b$ and substituting into Eq.(\ref{rt})
we obtain
\be\label{t}
t=-2i\sin k_1 e^{-ik_2L}(ae^{-ik_1}+b-e^{ik_2}(ce^{-ik_1}+d))^{-1}
\ee
\be\label{r}
r=-\frac{ae^{ik_1}+b-e^{ik_2}(ce^{ik_1}+d)}
        {ae^{-ik_1}+b-e^{ik_2}(ce^{-ik_1}+d)}
\ee
and
\be\label{Lan}
R_L(k_1,k_2)=(4\sin^2k_1)^{-1}|ae^{ik_1}+b-ce^{i(k_1+k_2)}-de^{ik_2}|^2
\ee
Recall that $R_L$ depends on $E$ via $a,b,c,d$ which are real
polynomials of $E$.
The ordinary choice for the potential outside the sample is $\o_1=\o_2=0$.
With this we get $k_1=k_2=k$,
$2\cos k=E$ and can `test the sample' at energies between $-2$ and $2$.

A nice and simple form is obtained for $k_1=k_2=\pi/2$ which
corresponds to $\o_1=\o_2=E$:
\[R_L(\pi/2,\pi/2)=|ia+b+c-id|^2/4=(\para T_{1\goto L}\para^2_2-2)/4\]

A truly intrinsic resistance is the minimum of $R_L$ over all $k_1,k_2$
which, in general, cannot be given in a closed form.
The minimum over $k_2$, when $k_1$ is fixed at $\pi/2$, is still easy to find:
\[\min_{k_2}R_L(\pi/2,k_2)=R_L(\pi/2,\mbox{arg}\,\frac{a-ib}{c-id})
=(1/4)(\sqrt{a^2+b^2}-\sqrt{c^2+d^2})^2\]
If the Lyapunov exponent is positive, the asymptotic behavior of the resistance
is the same for all $k_1,k_2$: $R_L$ grows with $L$ like $\exp\{2\g_+L\}$.
\newsec{\Sch equation with periodic potentials}
Periodic differential equations have a fully developed beautiful theory
(Floquet theory, see
e.g. [45]) which can be adapted without any difficulty to the
case of difference equations. Moreover, strictly ergodic sequences are well
approximated by periodic ones, and most of the results discussed in Sections
6 and 7 deeply rely upon this fact.
In this section we therefore survey the periodic case.

Let the potential be periodic with period $L\geq1$ finite:
\[V_{n+L}=V_n\ \ \mbox{for any integer}\ \ n\]
Consequently,
\[T_{m+L\goto n+L}=T_{m\goton}\ \ ,\ \mbox{all}\ \ m,n\]
We exploit this by writing $n=kL+m$ where $k=k(n)$ and
$0\leq m=m(n)<L$ are uniquely determined and, e.g. for $n>0$,
\[\P_n=\start{c}\p_{n+1}\\ \p_n\fin=T_{1\goton}\P_0=T_{1\goto m}(T_{1\goto L})^k
\P_0\]
According to the value of $E$ we can distinguish several cases (see also
Figure 1):\\
(1) The transfer matrix is elliptic or hyperbolic,
\[|\tr T_{1\goto L}(E)|=|\l+\l^{-1}|\neq 2\]
where $\l$ and $\l^{-1}$ are the eigenvalues. In this case there exists
a matrix $S$ (generally nonunitary) such that
\[S^{-1}T_{1\goto L}S=\start{cc}\l&0\\0&\l^{-1}\fin\]
Therefore we can write
\[\P_n=T_{1\goto m}S\start{cc}\l^k&0\\0&\l^{-k}\fin S^{-1}\P_0\]
The initial condition
\[\P_0=S\start{c}1\\0\fin\]
yields the solution
\[\F^1_n=\l^{k(n)}T_{1\goto m(n)}S\start{c}1\\0\fin\]
while with
\[\P_0=S\start{c}0\\1\fin\]
we get
\[\F^2_n=\l^{-k(n)}T_{1\goto m(n)}S\start{c}0\\1\fin\]
Because $m(n)$ is an $L$-periodic function, $\l^{\pm k}$ in $\F^{1,2}$
multiplies an $L$-periodic function of $n$. There are two cases:\\
(1.1) The transfer matrix is elliptic,
\[ \l=e^{i\eta}\ \ ,\ \ \tr T_{1\goto L}(E)=2\cos\eta\ \ ,\ \ \eta\neq
\mbox{integer}\,\times\pi\]
The two solutions $\f^{1,2}$ are Bloch waves and $E$ is in one of the
(at most $L$) `stability intervals'.
The Lyapunov exponent exists, $\g_+=\g_-=\g(E)=0$. The Landauer resistance
$R_n$ is a bounded quasiperiodic function of $n$ (periodic, if $\eta$ is a
rational multiple of $\pi$). \\
(1.2) The transfer matrix is hyperbolic, say, $|\l|>1$.
The two solutions can be written as
\[\f^1_n=\l^{\lfloor n/L\rfloor}p_n\ \ ,\ \ \f^2_n=\l^{-\lfloor n/L\rfloor}q_n\]
with $L$-periodic sequences $p_n,q_n$. The energy is in an open - maybe
semi-infinite - interval. The finite intervals are called `forbidden gaps', they
separate the stability intervals.
The Lyapunov exponent $\g(E)=L^{-1}\ln|\l|$. The Landauer resistance $R_n$ grows
exponentially (with exponent $2\g$).\\
(2) The transfer matrix is parabolic. There exists a unitary matrix $U$ such
that
\[U^{-1}T_{1\goto L}U=\start{cc}1&b\\0&1\fin\ \ ,\ \ b\neq0\]
The solution can therefore be written as
\[\P_n=T_{1\goto m}U\start{cc}1&kb\\0&1\fin U^{-1}\P_0\]
The initial condition
\[\P_0=U\start{c}1\\0\fin\]
gives rise to the periodic solution
\[\F_n=T_{1\goto m(n)}U\start{c}0\\1\fin\]
and any other solution oscillates and grows linearly with a slope $b/L$.
The energy is a boundary point of a stability interval. The Lyapunov exponent
is zero, the Landauer resistance has a quadratic increase.\\
(3) Accidentally it may happen that $T_{1\goto L}(E)=\pm I$. Then
the general solution
\[\P_n=T_{1\goto m(n)}\P_0\]
is periodic for all initial conditions. This situation is quoted as
the `closure of a forbidden gap' (cf. Figure 1b). In most concrete cases,
it is difficult
to say whether it occurs or not. If all the gaps are open, there are $L-1$
of them, separating $L$ disjoint stability intervals. Some examples :

(i) For the Almost-Mathieu equation with $\a=p/q$
rational and $q\o/\pi$ noninteger
all the ($q-1$) gaps are open. If $\o=0$, $p$ is odd and $q=4m$, then one gap
is closed (Bellissard, Simon [21], Choi, Elliot, Yui [31]).

(ii) For circle potentials (\ref{Sturm}) with rational $\a$ all the
gaps are open.

(iii) We can obtain periodic potentials by iterating a substitution
on a letter a finite
number of times and repeating the result periodically. Doing this with the
period-doubling substitution ($\s(a)=ab$, $\s(b)=aa$),
for the resulting equation
all the gaps are open (Bellissard, Bovier, Ghez [15],
[27]). For the Thue-Morse substitution ($\s(a)=ab$,
$\s(b)=ba$) the trace map ([9])
\[ x_{n+2}-2=(x_{n+1}-2)x_{n}^2\qquad(n\geq1)\]
with $x_n=\tr T_{1\goto 2^n}$ reveals that $x_n$ is a $2^n$ degree
polynomial of $E$, and $x_{n+2}-2$ has double zeros at the $2^n$ simple
zeros of $x_n$. This shows (see Fig. 1)
that at least one fourth of the gaps is closed! In fact,
$1/4$ is the precise proportion of the closed gaps
([12]; see also [10] for the corresponding phonon problem).

(iv) For the best periodic approximations of a hierarchical potential all
the gaps are open (Kunz, Livi, S\"ut\H{o} [90]).

We close this section with the conclusion that the physically relevant
solutions belong to energies which fall into the stability intervals:
\[|\tr T_{1\goto L}(E)|< 2\]
These solutions are the well-known Bloch waves.
The stability intervals are nothing else than the bands in the spectrum of the
periodic \Sch operator.
\newsec{Spectral theory}
Let us come back to the question we asked already in the previous
sections. The \Sch equation on $\Rr^d$ or $\Zz^d$ can be solved for any $E$.
Which are the physically relevant solutions?
What can be said about the corresponding energies?
Spectral theory provides the following answer:\\
(i) To compute matrix elements of functions of the energy operator, it
suffices to consider only
the polynomially bounded solutions of the \Sch equation.\\
(ii) Every subexponential solution (growing slower than any exponential)
belongs to an energy which is in the
spectrum of the energy operator.
\subsection{\Sch operator, $\ell^2$-space and spectrum}
Define an operator $H$ on the sequences $\p=\{\p_n\}$ by
\[(H\p)_n=\p_{n-1}+\p_{n+1}+V_n\p_n\]
Then Eq.(\ref{Sch}) is equivalent to $H\p=E\p$. We call $H$ the \Sch operator;
it can be represented with the infinite tridiagonal (Jacobi)
matrix (denoted by the same letter)
\be\label{Jacobi}H=\start{cccccc}\cdot&\cdot&\cdot&\ &\ &\ \\
\ &1&V_{n-1}&1&\ &\ \\
\ &\ &1&V_n&1&\ \\
\ &\ &\ &\cdot&\cdot&\cdot\fin\ee
Consider
\[\eltu=\left\{\p:\suminf |\p_n|^2<\infty\right\}\]
This is a {\em separable} Hilbert space (it contains a countable dense set)
with the inner product $(\vf,\p)=\suminf\vf^*_n\p_n$,
and $H$ is a selfadjoint operator on it.  In what follows, we use the
notation $\hilb$ for a separable Hilbert space. We call the elements of $\hilb$
{\em vectors}. The norm of a vector $\p$ is defined by $\|\p\|^2=(\p,\p)$.
It should not be confounded with the norm of the $\Cc^2$ vectors which we
introduced in
Eq.(\ref{vector}) and denote by capital Greek letters.

Suppose, for the sake of simplicity, that $V$ is a bounded sequence.
{\em The spectrum} $\s(H)$
of $H$ is the set of $E$ values for which $E-H$ has no
bounded inverse: One can find some vector $\p$ such that
the equation
\be\label{inhom}
(E-H)\vf=\p
\ee
has no vector solution for
$\vf$ (but may have a solution outside $\hilb$, see Problem 5.2.2) or it has
more than one vector solutions. In the second case
$(E-H)\vf=0$ has a nontrivial vector
solution, that we call an {\em eigenvector} of $H$, belonging to the
{\em eigenvalue} $E$.
(Recall that the spectrum of a finite matrix $A$ is the set of its
eigenvalues, and
\[\det(\l-A)=0\ \mbox{if and only if}\ (\l-A)^{-1}\ \mbox{does not exist.})\]

The spectrum is real, nonempty and closed, i.e., contains all of its
limit points.

{\em Example.} Let us decompose $H$ into off-diagonal and diagonal parts,
$H=H_0+V$. Then $\s(H_0)=[-2,2]$. If $V$ is L-periodic,
\[\s(H)=\{E: |\tr T_{1\goto L}(E)|\leq 2\}\]
In either cases, the \Sch equation has no vector solution, thus there is
no eigenvalue in the spectrum.

If $E$ is an isolated point in $\s(H)$, then $E$ is an eigenvalue.

The {\em essential spectrum}, $\sess(H)$, of $H$ is the set of all
non-isolated points of $\s(H)$, together with the isolated eigenvalues of
infinite multiplicity (do not occur in one dimension). Clearly, this is a
closed set. An important
characterization of the essential spectrum is due to Weyl: \\
$E$ is in
$\sess(H)$ if and only if it is an {\em approximate eigenvalue} of $H$
in the following sense. One can find a sequence $\vf^n$ of
vectors such that $\para\vf^n\para=1$, $\vf^n$ goes to zero weakly
(in $\eltu$ this means that $\vf^n_k\goto 0$ for each $k$ fixed:
either $\vf^n$ spreads over larger and larger domains or it remains well
localized but `escapes' to infinity) and
\be\label{Weyl}H\vf^n=E\vf^n+\ve^n\ee
where $\ve^n$ are vectors, $\para\ve^n\para\goto 0$.

If $\p$ is not an eigenvector but it is a subexponential solution of
$H\p=E\p$, one can
construct such a `Weyl sequence', so $E\in\sess(H)$.
If $\p$ is polynomially bounded, we call it a {\em generalized
eigenvector (eigenfunction)} of $H$, and $E$ a {\em generalized eigenvalue}.

{\em Problem 1.} Construct a Weyl sequence for a subexponential solution $\p$
which is not an eigenvector.
{\em Hint:} Replace $\p_k$ by $0$ for $k<0$ and $k>n$ and normalize.
\subsection{Point spectrum}
The point spectrum of $H$ is the set of all eigenvalues of $H$. It is denoted
by $\spp(H)$. The point spectrum may be empty, finite or countably
infinite. The {\em pure point subspace} of $\hilb$ is the subspace spanned by
all the eigenvectors of $H$. It is denoted by $\hpp$.

We say that $H$ has a {\em pure point spectrum}, if its eigenvectors form a
basis in the Hilbert space, that is, $\hilb=\hpp$. In this case the spectrum
is the closure of the set of eigenvalues:
\[\s(H)=\overline{\spp(H)}=\spp(H)\cup\{\mbox{limit points of $\spp(H)$}\}\]

{\em Example.} A nontrivial
counterexample known for every physicist is the hydrogen atom.
Here the Hilbert space is $L^2$($\Bbb R$$^3$). The energy operator of the
hydrogen atom has infinitely many bound states (eigenvectors)
which, however, do not span $\hilb$. So $\hpp$ is an actual part
of $\hilb$. $\spp$ is a discrete set of negative numbers with a unique
accumulation point, $0$.
Any $E>0$ belongs to the spectrum but {\em is not} an eigenvalue,
only a generalized eigenvalue. The corresponding generalized eigenvectors are
bounded: they are
products of a plane wave with a confluent hypergeometric function. The
generalized eigenvectors (`scattering waves') can be used to build up
wave packets, i.e., square integrable functions (vectors of
$\hilb$) which are orthogonal to every eigenvector.

{\em Problem 1.} Find a possible Weyl sequence to show explicitly that $0$ is
in the essential spectrum of the energy operator of the hydrogen atom.

$\spp$ is called a {\em dense point spectrum} if it is nonempty and has no
isolated point. If, for a bounded recurrent potential, there is a point
spectrum, it is a dense point spectrum (see Problem 6.2.2).\\
{\bf Examples of dense pure point spectra}\\
1. Let $H=V$, i.e., a diagonal matrix with $V_n=\cos2\pi n\a$, where $\a$ is
irrational. $V$ has a pure point spectrum, dense in $[-1,1]$:
\[\spp(V)=\{V_n\}_{n=0}^{\infty}\ \ ,\ \ \s(V)=[-1,1]\]
The eigenvector belonging to the eigenvalue $V_n$ is $\d^n$, the unit vector
concentrated on the site $n$. The set of eigenvectors is the canonical basis in
$\eltu$.

{\em Problem 2.} Choose $E$ in $[-1,1]$ but $E\neq V_n$, any $n$. Use
the definition of the spectrum to show that $E\in\s(V)$. \\
2. Anderson localization. Let $V_n$ be identically distributed
independent random variables. Let $V(\o)=\{V(\o)_n\}$ denote a realization,
$H(\o)=H_0+V(\o)$.

(i) Pastur [106]: There exists a set $\S\subset$$\Bbb R$ such that
\begin{itemize}
\item $\s(H(\o))=\S$ with probability $1$.
\item $\S$ contains no isolated points.
\item Any $E\in\S$ is {\em not} an eigenvalue with probability $1$.
\end{itemize}

(ii) Kunz, Souillard [91](simplified): Let the probability distribution
$r(x)$ of the $V_n$'s be continuous, and $r(x)\neq0$ if and only if $a<x<b$.
Then
\begin{itemize}
\item the spectrum of $H(\o)$ is pure point with probability $1$.
\item $\S=[a-2,b+2]$
\item All the eigenvectors are exponentially localized.
\end{itemize}

(iii) Carmona, Klein, Martinelli [28]: Let $V_n$ be
Bernoulli-distributed,
\[V_n=\opt 0& \mbox{with probability $p$}\\b&\mbox{with probability $1-p$}
\finopt\]
Then the spectrum is pure point with probability $1$,
\[\S=[-2,2]\cup[b-2,b+2]\]
and all the eigenvectors are exponentially localized.\\
3. Almost Mathieu equation ($\l>0$).

Choose $\o$ at random in the interval $[0,1)$ (according to the uniform
distribution). There is an appealing nonrigorous argument (Aubry-Andr\'e
duality, [4]), according to which the spectrum should be pure
point for $\l>2$. This is not quite true, but seems to hold for good Diophantine
$\a$ and almost every $\o$:

(i) If $\a$ is irrational, the spectrum is independent of $\o$,
$\s(H(\o))=\S(\a,\l)$.

(ii) Sinai [131], Fr\"ohlich, Spencer and Wittwer [50]: If $\a$
is a good Diophantine number, namely, there exists some constant $c>0$
such that
\be\label{Dioph}
\min\{n\a-\lfloor n\a\rfloor,\lceil n\a\rceil-n\a\}\geq c/n^2
\ \ \mbox{for all}\ \ n\neq0
\ee
and $\l\gg1$, the spectrum of $H(\o)$ is pure point for almost every $\o$
and the eigenvectors decay exponentially. Sinai's proof yields also that
$\S(\a,\l)$ is a Cantor set (see below).\\
Aubry, Andr\'e [4], Thouless [138], [139]: The
Lebesgue measure $m$ of the spectrum is positive,
\[m(\S(\a,\l))\geq2|\l-2|\ ,\ \frac{m(\S(\a,\l))}{E_u-E_l}\geq
\frac{\l-2}{\l+2}\]
($E_u$ and $E_l$ are the upper and lower boundary of the spectrum,
respectively.) Recently it was shown by Last ([93], [94])
that for good Diophantine $\a$, as for example in (\ref{Dioph}),
\[m(\S(\a,\l))=2|\l-2|\]

(iii) Jitomirskaya [69], [70]  brought sensible
improvements to the above results. She simplified the proof, weakened the
Diophantine condition on $\a$, put an explicit bound on $\l$ ($\l>15$), and
showed that for all $\l>2$ and for almost every $\o$ the closure of the point
spectrum has the same Lebesgue measure as the spectrum itself.

(iv) One may think that there is only some technical difficulty to
extend the above results on localization from a.e $\o$ to {\em every} $\o$.
This is not true: Recently, Jitomirskaya and Simon [71] proved that for
any $\l>2$ and irrational $\a$ there is an uncountable set of $\o$'s
which is dense in the interval $[0,1]$ (i.e., any point in $[0,1]$ is a
limit point of this set) and for which the spectrum of $H(\o)$ is purely
singular continuous (see later). Clearly, this set is of zero Lebesgue
measure, if for the given $\l$ and $\a$ localization occurs for a.e. $\o$.
This result provides an example of a singular continuous measure the support
of which is a thick (positive-measured) Cantor set.
\subsection{Cantor sets}
Take a closed interval $C_0$ of the real line. Cut off a finite number
of open intervals which have no common boundary points with each other and
with $C_0$. What remains is a closed set $C_1$ without isolated points:
a union of a finite number
of closed intervals. Repeat the same procedure with $C_1$ to obtain $C_2$,
and so on: continue it indefinitely, by following the rule that no interval
is left untouched. Let $C$ denote the resulting set. It has the following
properties:
\begin{enumerate}
\item $C$ is closed. Indeed, each $C_n$ is closed,
$C_0\supset C_1\supset\cdots$,
therefore
$C=\cap_{n=0}^{\infty}C_n$,
and any intersection of closed sets is closed.
\item $C$ is nonempty. Clearly, the boundary points $c_{ni}$ of $C_n$ are
not removed during the construction, therefore $C$ contains them for all $n$.
\item There can be no interval in $C$, according to the construction rule.
\item There is no isolated point in $C$. In fact, it is clear from the
construction, that the set of points $c_{ni}$ is dense in $C$.
\end{enumerate}
A set with the above four properties is called a Cantor set.
The boundary points of $C^c$, the complement of the Cantor set $C$,
form a countable dense set in $C$, but the Cantor set itself
is uncountable. $C^c$ is an infinite union of open intervals,
the closure of which is the whole real line. What is the
Lebesgue measure of $C$? Suppose that $m(C_{n+1})=x_nm(C_n)$, then
\[m(C)=m(C_0)\prod_{n=0}^{\infty}x_n\]
This number is greater than zero if and only if
\[\sum_{n=0}^{\infty}\ln x_n>-\infty\]
which is true if and only if
\[\sum_{n=0}^{\infty}(1-x_n)<\infty\]
For Cantor's `middle thirds' set $x_n=2/3$ and thus the Lebesgue measure
is zero.
\subsection{Continuous spectrum}
Let $\hcont=\hpp^{\perp}$, the subspace orthogonal to
$\hpp$. By definition, $H$ has no eigenvector in $\hcont$.
If the only element of $\hcont$ is the null vector, the spectrum is
pure point. If $\hcont$ contains a nonzero vector, $H$ has (also) a continuous
spectrum $\scont(H)$. Indeed, the restriction of $H$ to $\hcont$,
denoted by $H|\hcont$,  is a
selfadjoint operator whose spectrum is nonempty, and
\[\scont(H)=\s(H|\hcont)\]
As a matter of fact, the
dimension of $\hcont$ is either $0$ or infinite
($\hcont$ is an $H$-invariant subspace ($H\hcont\subset\hcont$), and $H$
has eigenvectors in any finite dimensional invariant subspace)
and in the second case $\scont(H)$ is a nonempty closed
set without isolated points. Thus we have the decomposition
\[\hilb=\hpp\oplus\hcont \qquad,\qquad \s(H)=\overline{\spp(H)}\cup\scont(H)\]
but notice that the closure of $\spp$ (even $\spp$ itself)
may overlap with $\scont$.

%

%
Just as to the points of $\spp(H)$ there correspond the $H$-eigensubspaces of
$\hpp$, to certain uncountable subsets of $\scont(H)$ there correspond
infinite-dimensional $H$-invariant subspaces of $\hcont$. The connection
is made by spectral projections.
\subsection{Spectral projections}
If $H$ has a pure point spectrum, it can be decomposed as
\[H=\sum_{E\in\spp(H)}E\pe\]
where $\pe$ is the orthogonal projection to the eigensubspace belonging to
the eigenvalue $E$ (for the one dimensional \Sch operator all these
subspaces are one dimensional). The orthogonal projections are selfadjoint
and idempotent,
\[\pe^{\dagger}=\pe=\pe^2\]
The orthogonality of the eigensubspaces belonging to different eigenvalues
can be expressed as
\[\pe P(\{E'\})=\d_{E,E'}\pe\]
and, if the spectrum is pure point,
\[\sum_{E\in\spp(H)}\pe=I=\,\mbox{identity}\]

{\em Physicists' notation:}\\
If $E$ is a nondegenerate eigenvalue, $\p$ the normalized eigenvector, then
\[\pe=|\p\rangle\langle\p|\]
If $E$ is degenerate,
\[\pe=\sum|\p^i\rangle\langle\p^i|\]
where one has to sum over an orthonormal basis in the eigensubspace.

In general, there is a fundamental relation between certain subsets of the
spectrum and orthogonal projections to $H$-invariant subspaces.
Let $\D=(a,b]$ be a half-open
interval of the
real line, $\hid$ its characteristic function, i.e., $\hid(E)=1$ if $E\in\D$
and $0$ otherwise. This function satisfies
\be\label{hid}
\hid(E)=\hid(E)^*=\hid(E)^2
\ee
One can define an operator $\hid(H)$ (formally obtained by substituting $E$
with $H$) as a strong limit (see below)
of polynomials of $H$, which inherits the properties
(\ref{hid})
of the real-valued function $\hid$ and is, therefore, an orthogonal projection;
we denote it by $\pdel$. Similarly, orthogonal projections can be defined
for more complicated sets, called Borel sets, as, for example,
countable unions and intersections of half-open intervals
(closed and open sets are Borel sets), and the family of the corresponding
projections have nice algebraic properties:
\[P(\emptyset)=0\ ,\ P(\Rr)=I\]
\[P(\D_1\cap\D_2)=P(\D_1)P(\D_2)\]
and for $\D_1$ and $\D_2$ disjoint sets
\[P(\D_1\cup\D_2)=P(\D_1)+P(\D_2)\]
The last equality holds also in a stronger form: If $\D_i$, $i=1,2,...$ is an
infinite sequence of pairwise disjoint Borel sets then
\[P(\cup_{i=1}^\infty\D_i)=\mbox{s-}\!\lim\sum_{i=1}^N P(\D_i)\]
The sum converges in the strong sense, that is, for any $\vf\in\hilb$
\[\sum_{i=1}^NP(\D_i)\vf\goto P(\cup_{i=1}^{\infty}\D_i)\vf\ \
\mbox{as}\ \ N\gotoinf\]
The above properties of $P$ are characteristic to probability measures; the only
difference is that $\pdel$ are operators, so $P$ is a projection-valued
probability measure.
This measure is called the {\em (spectral) resolution of the identity}, its
values on Borel sets are the {\em spectral projections}.

The spectral projections live on the spectrum of $H$:
\[\pdel=P(\D\cap\s(H))\]
If $\pdel\neq0$,
the linear subspace $\pdel\hilb$ is nontrivial and
invariant under $H$ (because $H$ commutes with $\pdel$).

Consider now the converse relation: Let $\subh$ be an $H$-invariant (closed)
subspace
and denote $[\subh]$ the orthogonal projection onto $\subh$. $[\subh]$ may
not be a spectral projection. For example, if $\p$ is an eigenvector belonging
to a degenerate eigenvalue $E$ and $\subh$ is the one-dimensional subspace
spanned by $\p$, $[\subh]=|\p\rangle\langle\p|$ is not a spectral projection.
The smallest spectral projection larger than $[\subh]$ is $\pe$ (for
projections $Q_1$, $Q_2$, $Q_1<Q_2$ means $Q_1\hilb\subset Q_2\hilb$).
Similar situation may occur in the continuous spectrum. The operator $H$ is
called {\em multiplicity free} if the orthogonal projection onto any
$H$-invariant subspace is a spectral projection. Clearly, $[\hpp]$ and
$[\hcont]$ are always spectral projections:
\[ [\hpp]=P(\spp(H))\ \ \mbox{and}\ \ [\hcont]=P(\s(H)\setminus\spp(H))\]
Further important examples will be given later.

 For  any vector $\p$ one can define a real positive measure $\mup$ by setting
 \[\mup(\D)=(\p,\pdel\p)=\para\pdel\p\para^2\]
 for Borel sets $\D$. This is called the {\em spectral measure associated with
 $\p$}. Spectral measures play an important role in the computation of averages
 and transition amplitudes: The matrix elements of functions of the energy
 operator can be obtained as integrals with respect to these measures.
The spectral projection
\[P_E=P((-\infty,E])\]
is a monotonically increasing function of $E$ in
the sense that for $E_1<E_2$
\[P_{E_2}-P_{E_1}=P((E_1,E_2])\geq0\]
($P\geq0$ means $(\p,P\p)\geq0$ for all $\p\in\hilb$, which holds because
$(\p,P\p)=(P\p,P\p)$.) Therefore, with
\[\dd P_E=P_{E+\dd E}-P_E=P((E,E+\dd E])\]
one can write down the spectral decomposition of $H$ in the
general case:
\[H=\int E\dd P_E\]
This equation has the following meaning.
For any $\p\in\hilb$ and any function $f$,
continuous on $\s(H)$,
\[(\p,f(H)\p)=\int f(E)\dd\m_{\p}(E)\]
where
\[\mpe=\mup((-\infty,E])=\para P_E\p\para^2 \]
This function increases monotonically with $E$ and is upper semicontinuous
(in jumps takes on the higher value). It fully determines the spectral measure
$\mup$.
In the above equation the integration is done with this measure.

Here we open a parenthesis on measures. The main line of the discussion
continues in Section 5.8.
\subsection{Measures}
Let $\m(x)$ be a real monotonically increasing upper semicontinuous
function on $\Bbb R$, finite at any finite $x$. With this function is
associated a measure on the Borel sets $\D$ of $\Bbb R$:
\[\m(\D):=\int_{\D}\dd\m(x)=\int\hid(x)\dd\m(x)\]
which is nothing else than the total variation (increase) of $\m$ on $\D$,
written as the Stieltjes-Lebesgue integral of $\hid(x)$ with respect to $\m(x)$.
(The notation should not confound the reader: $\m$ with a real number argument
means the function, with a set argument the measure. In this way,
$\m(x)=\m((-\infty,x])$ and $\m(\{x\})$ is the measure of the one-point set
$\{x\}$.)
The set function thus obtained satisfies, indeed, the properties of
(positive) measures, namely, it is nonnegative, vanishing on the empty set
and countably additive:
\[\m(\cup_{i=1}^{\infty}\D_i)=\sum_{i=1}^{\infty}\m(\D_i)\]
if the $\D_i$ are pairwise disjoint sets.

Conversely, a positive measure $\m$ which is finite on the semi-infinite
intervals $(-\infty,x]$ can be used to define a monotonically increasing upper
semicontinuous function, denoted also by $\m$, by setting
\[\m(x):=\m((-\infty,x])\]
This is called the {\em distribution function} of the measure $\m$.

One can define the derivative of the measure $\m$ with respect to the
Lebesgue measure at the point $x$ as
\[\frac{\dd\m}{\dd m}(x)=\lim_{J\downarrow x}\m(J)/m(J)\leq\infty\]
provided the limit exists.
The limit is taken on open intervals containing $x$ and shrinking to $x$.
If the distribution function is continuous and differentiable at $x$ (perhaps
with infinite derivative) then its derivative
$\m'(x)=\mpx$. Oppositely, if $\mpx$ {\em finitely}
exists then $\m(x)$ is continuous and differentiable at $x$ and $\m'(x)=\mpx$.
In the points of discontinuity of $\m(x)$, $\m'(x)$ does not
exist, but $\mpx$ exists and is infinite. The set
\[S=\{x: \mpx\ \mbox{does not exist finitely or infinitely}\}\]
is a part of the set where $\m'(x)$ does not exist. Therefore we can apply
a theorem ([124], Ch.~IV, Theorem 9.1),
valid for functions which are locally of bounded variation
(like $\m(x)$), to conclude that both the $\m$ and the Lebesgue ($m$-) measure
of $S$ is vanishing, so $\mpx$ is well-defined apart from a set of zero
$\m$- and Lebesgue measure.

Let $A$ be a Borel set.
We say that a measure $\m$ {\em is concentrated on} $A$ if $\m(A^c)=~0$
($A^c=\Rr\setminus A$, the complement of $A$).
The smallest closed set on which
$\m$ is concentrated is the {\em support} of $\m$, $\suppm$. So if $\m$ is
concentrated on $A$ then $\suppm\subset\overline{A}$, the closure of $A$.

An $x\in\Rr$ is a {\em point of increase} of the distribution function $\m$
if for every $\ve>0$, $\m(x+\ve)>\m(x-\ve)$.
\medskip

{\em Problem 1.} $\suppm$ is the set of points of increase of $\m(x)$.
\medskip

A Borel set $A$ is called
an {\em essential support} of $\m$ if $\m(A^c)=0$ and for any $A_0\subset A$
such that $\m(A_0)=0$, $m(A_0)=0$ as well. So the support is unique and closed,
the essential support is unique only apart from sets of zero $\m$- and
Lebesgue measure and is normally not closed.
Countable unions and intersections of essential supports are
essential supports.
An essential support may not be a subset of the
support. An example is $A\cup B$ where $A$ is an essential support and $B$ a
nonempty Borel set with $m(B)=0$ and $B\cap\suppm=\emptyset$. However, if $A$
is an essential support then $A\cap\suppm$ is again an essential support,
the closure of which is the support.

It follows from the definitions, that any measure $\m$ has an essential support
on which $\mpx$ exists (maybe infinite).

Any measure $\m$ can be decomposed into pure point and continuous part,
$\m=\mpp+\mcont$. These are respectively characterized by the equations
\[\mpp(\Rr)=\sum_i\mpp(\{x_i\})\]
that is, $\mpp$ is concentrated on a countable set of points, and
\[\mcont(\{x\})=\lim_{\ve\downarrow0}\m([x-\ve,x+\ve])=0,\]
the $\mcont$-measure of every $x\in\Rr$ is zero. If the function $\m$ is
continuous, the corresponding measure is purely continuous; if it
makes jumps of heights $a_i$ in the points $x_i$,
\be\label{mpp}
\dd\mpp(x)=\sum_ia_i\d(x-x_i)\dd x
\ee
For $\mpp$ there exists a smallest essential support:
the set of all points of discontinuity of the function $\m$.

If $\m$ and $\n$ are two measures, we say that $\m$ is {\em absolutely
continuous} with respect to $\n$, and write $\m\ll\n$, if $\m(\D)=0$
whenever $\n(\D)=0$. In words, if $\n$ is concentrated on a set, then $\m$ is
concentrated on the same set.
If $\m$ and $\n$ vanish on the same Borel sets, i.e., they are
mutually absolutely continuous, we call them {\em equivalent}, and write
$\m\sim\n$. Equivalent measures have the same support and essential supports.
Most often $\n=m$, the Lebesgue measure. If we say only
that $\m$ is absolutely continuous, we mean it with respect to the Lebesgue
measure. Clearly, if $\m$ is absolutely continuous, it is continuous.

{\em Problem 2.} If two measures are absolutely continuous and have a common
essential support then they are equivalent.

{\em Problem 3.} Find an absolutely continuous and a pure point
measure whose supports are not essential supports. Find a measure,
whose support is the smallest essential support.

A function $f$ defined on $\Rr$ is said to be {\em absolutely
continuous} on a finite or infinite interval $J$ if to every $\ve>0$
there exists a $\d>0$ such that for any finite collection of disjoint open
subintervals $(a_1,b_1),(a_2,b_2),...,(a_n,b_n)$ of $J$
\be\label{abscont}
\sum_i|f(b_i)-f(a_i)|<\ve\ \ \mbox{whenever}\ \ \sum_i(b_i-a_i)<\d
\ee
Since $\d$ does not depend on $n$, the property holds also if
the number of intervals is infinite.

{\em Problem 4.} A measure is absolutely continuous if and
only if its distribution function is absolutely continuous on every finite
interval.

We say that two measures $\m$ and $\n$ are {\em mutually singular}, and
write $\m\perp\n$, if there are disjoint Borel sets $A$ and $B$ such that $\m$
is concentrated on $A$ and $\n$ is concentrated on $B$. Most often $\n=m$
and we simply say that $\m$ is singular. Any pure point measure is obviously
singular.

{\em Problem 5.} Any essential support of a singular measure is of zero
Lebesgue measure.

A measure is said to be {\em singular continuous} (with respect to the
Lebesgue measure) if it is singular and continuous. An example is given in the
next subsection.
Thus, $\m$ has the following decompositions into pairwise
mutually singular measures:
\[\m=\mpp+\mcont=\mac+\msing=\mac+\msc+\mpp\]

{\em Problem 6.} Show that any measure uniquely determines its {\em pp, ac}
and {\em sc} parts.
\medskip

According to the value of $\mpx$, one can find essential
supports for the different components as follows (taken over from
Gilbert and Pearson [52]). Let
\[B=\{x\in\Rr: \mpx\ \mbox{exists,}\ 0\leq\mpx\leq\infty\}\]
The sets $M,\Mac,\Msing,\Msc,\Mpp$ defined below are essential supports
respectively for $\m,\mac,\msing,\msc,\mpp$:
\beast\begin{array}{lll}
 M    &=&\{\xinb 0<\mpx\leq\infty\}\\
\Mac  &=&\{\xinb 0<\mpx=\m'(x)<\infty\}\\
\Msing&=&\{\xinb \mpx=\infty\}\\
\Msc  &=&\{\xinb \mpx=\infty,\ \m(\{x\})=0\}\\
\Mpp  &=&\{\xinb \mpx=\infty,\ \m(\{x\})>0\}
\end{array}
\east
Obviously, $\Mac,\Msc$ and $\Mpp$ are pairwise disjoint sets. They are subsets
of $\suppm$ and $\Mpp\subset\supp\mpp$ ($\Mpp$ is the smallest essential
support of $\mpp$), but $\Mac$, $\Msing$ and $\Msc$ may not be parts of the
supports of the respective measures. As mentioned earlier, by intersecting them
with the corresponding supports, we still get essential supports. It may happen,
however, that $\Mac$ is nonempty but $\mac=0$: surely this is the case if
$m(\Mac)=0$; or $\Msc$ is nonempty, even uncountable, but $\msc=0$. For more
details see [52].
\subsection{Cantor function}
The Cantor function is a continuous monotonically increasing function which
grows from $0$ to $1$ exclusively in the points of the middle-thirds
Cantor set
\[C=\{\sum_{n=1}^{\infty}3^{-n}x_n:\ x_n=0\ \mbox{or}\ 2\}\]
In the interval $[0,1]$ it is given by the formula
\[\a(x)=\sum_{n=1}^{\infty}2^{-n-1}x_n\ \ \mbox{if}\ \
x=\sum_{n=1}^{\infty}3^{-n}x_n\ \ ,\ \ x_n\in\{0,1,2\}\]
with the remark that triadic rationals are taken with their infinite
representation. More interesting, it is entirely determined by the conditions
\begin{enumerate}
\item $2\a(x)=\a(3x)\ ,\ 0\leq x\leq1/3$
\item $\a(x)+\a(1-x)=1$
\item $\a(x)$ is monotonically increasing
\item $\a(x)=0$ if $x<0$ and $\a(x)=1$ if $x>1$
\end{enumerate}
from which the moments or the Fourier transform of $\a$ can be computed. For
example, the latter is
\[\int e^{itx}\dd\a(x)=e^{it/2}\prod_{n=1}^{\infty}\cos t/3^n\]
The function is represented by a `devil's staircase'.
The measure $\a$ is singular continuous, because it is continuous and
its support, $C$, is of zero Lebesgue measure. Observe that the Fourier
transform does
not tend to zero when $|t|$ goes to infinity (check it with
$t=2\pi 3^k$) as it would do if $\a$ were absolutely continuous.

The support of a singular continuous measure may have positive Lebesgue measure.
Define, for example, an odd function ($\m(-x)=-\m(x)$) on $\Rr$ by setting
\[\m(x)=\sum_{q=1}^{\infty}2^{-q}\sum_{p=0}^{\infty}2^{-p}\a(x-p/q)\qquad
\mbox{if $x\geq0$}\]
This is a continuous function increasing in every point of $\Rr$:
$\m(x+\ve)-\m(x)>0$ for every real $x$ and positive $\ve$. $\m(x)$
is therefore the distribution function of a continuous measure $\m$
whose support is $\Rr$. Furthermore, $\cup_{r\in\Qq}(C+r)$ is a Borel set
(a countable union of closed sets) of zero Lebesgue measure
and is an essential support for $\m$; thus, $\m$ is singular continuous.
This situation may occur for \Sch operators with unbounded potentials. If
$V_n=\tan n\a$ and $\a$ is a Liouville number (an irrational number which is
extremely well approximated by rationals), all the spectral measures are
purely singular continuous, and there are spectral measures the support of which
is $\Rr$ ([130]).

{\em Problem 1.} Give an example of a pure point and a singular continuous
measure having a common essential support.
\subsection{Spectral measures and spectral types}
In the comparison of two measures, $\m$ and $\n$, yielding $\m\ll\n$,
$\m\sim\n$ or $\m\perp\n$, either or both can be projection-valued, and
a projection-valued measure has support, essential support and uniquely
determined {\em pp, ac} and {\em sc} parts, like real measures. In particular,
for any vector $\p$, $\mup\ll P$, where $P$ is the spectral resolution of the
identity. Similar relations for the {\em pp, ac, sc} parts will be written
down below. Let us start with the decomposition of the spectral resolution
of the identity:
\[P=\Ppp+\Pcont=\Pac+\Psing=\Ppp+\Pac+\Psc\]
Is it true that the different terms are {\em spectral} projection-valued
measures? The answer is yes, but for the moment
it is not even clear that $\Pac(\D)$ and $\Psc(\D)$
project onto $H$-invariant subspaces.

By definition,
\[\Ppp(\D)=\sum_{E\in\D\cap\spp(H)}\pe=P(\D\cap\spp(H))\]
which is a spectral projection. The smallest essential support of $\Ppp$
is $\spp(H)$. Notice that
\[\Ppp(\D)=P(\spp(H))\pdel=[\hpp]\pdel\]
where the two projections commute. Let $\p\in\hpp$. Then
\[\mup(\Rr)=\|\p\|^2=\|\Ppp(\Rr)\p\|^2=\sum_{E\in\spp(H)}\mup(\{E\})\]
i.e., $\mup$ is a pure point measure and $\mup\ll\Ppp$.

Also, by definition, for any Borel set $\D$,
\[\Pcont(\D)=\pdel-\sum_{E\in\D\cap\spp(H)}\pe=P(\D\setminus\spp(H))\]
is a spectral projection. The support of $\Pcont$ is $\scont(H)$. Notice that
\[\Pcont(\D)=(I-[\hpp])\pdel=[\hcont]\pdel=P(\spp(H)^c)\pdel\]
where the two projections commute. Let $\p\in\hcont$. Then for any $E\in\Rr$
\[\mup(\{E\})=\|\pe\p\|^2=\|\pe[\hcont]\p\|^2=\|\Pcont(\{E\})\p\|^2=0\]
i.e., $\mup$ is purely continuous and $\mup\ll\Pcont$.

A purely continuous spectral measure can still be decomposed into absolutely and
singular continuous parts. We would like
to perform the analogous decomposition on $\hcont$, $\scont(H)$ and
$\Pcont$.\vspace{3mm}

{\em Problems.}
\begin{enumerate}
\item $\mup$ is concentrated on a Borel set $A$ if and only if $P(A)\p=\p$.
      As a consequence, if $\p\in P(A)\hilb$, $\supp\mup\subset \overline{A}$.
\item $\m_{H\p}\ll\mup$. In particular, if $\mup$ is absolutely or
      singular continuous then $\m_{H\p}$ is absolutely or
      singular continuous, respectively.
      ({\em Remark.} We know already that if $\mup$ is pure point,
      $\m_{H\p}$ is also pure point
      because $\p\in\hpp$ and $H\hpp\subset\hpp$.)
\item If $\mup=(\mup)_\xpp+(\mup)_\xac+(\mup)_\xsc$
      then there exist orthogonal vectors $\ppp$, $\pac$ and $\psic$
      such that $\p=\ppp+\pac+\psic$ and $(\mup)_{\xpp}=\m_{\ppp}$,
      $(\mup)_{\xac}=\m_{\pac}$ and $(\mup)_{\xsc}=\m_{\psic}$.
      {\em Hint.} Choose disjoint essential supports and apply 1.
\item Let $\p^i$ be a sequence of vectors converging to a vector $\p$
(i.e., $\para \p^i-\p\para\goto0$). Suppose that $\m_{\p^i}$ are singular
measures. Then $\mup$ is singular. {\em Hint.} (i) $\m_{\p^i}\goto \mup$
on Borel sets. (ii) Let $A_i$ be an essential support to $\m_{\p^i}$,
then $\cup A_i$ is an essential support to $\mup$.
\end{enumerate}
It follows from Problem 2 that the vectors generating a purely {\em ac}
spectral measure form an $H$-invariant subspace $\hac$; similarly, the vectors
giving rise to a purely {\em sc} spectral measure form an $H$-invariant
subspace $\hsc$. (That the vectors of the same pure spectral type form a
{\em subspace} is the subject of Problem 7 below.)
According to Problem 3, $\hac$ and $\hsc$ are
orthogonal and span $\hcont$.
Thus $\hac$ and $\hsc$ are necessarily closed and
\[\hcont=\hac\oplus\hsc\]
(Problem 4 shows also explicitly that $\hsc$ is closed.)
Now we can consider the restrictions of $H$ to $\hac$ and to $\hsc$: these are
selfadjoint operators, their spectra are $\sac(H)$ and $\ssc(H)$,
respectively.
Absolutely continuous and singular continuous spectral measures are
concentrated respectively on $\sac(H)$ and $\ssc(H)$.

In summary, we obtained the following decompositions:
\[\hilb=\hpp\oplus\hac\oplus\hsc\]
Correspondingly, for any $\p\in\hilb$,
\[\p=\ppp+\pac+\psic\]
and
\[\mup=\m_{\ppp}+\m_{\pac}+\m_{\psic}\]

What about the decomposition of $\Pcont$? Because
\[[\hcont]=[\hac]+[\hsc]\]
where the two projections are orthogonal, we get
\[\Pcont(\D)=[\hcont]\pdel=[\hac]\pdel+[\hsc]\pdel\]
with commuting operators in the products.

{\em Problem 5.} Show that the projection-valued measures $[\hac]P$ and
$[\hsc]P$ are absolutely and singular continuous, respectively.

Since the decomposition of $\Pcont$ into {\em ac} and {\em sc} parts is unique,
we obtained that
\be\label{pacsc}
\Pac=[\hac]P\ \ ,\ \ \Psc=[\hsc]P
\ee
So $\Pac(\D)$ and $\Psc(\D)$ project onto $H$-invariant subspaces within
$\hac$ and $\hsc$, respectively. It remains to show that they are spectral
projections.

Let $A$ and $B$ be disjoint essential supports for $\Pac$ and $\Psc$,
respectively, which are disjoint also from $\spp(H)$. Then
\[[\hac]=\Pac(A)=P(A)\ \ \mbox{and}\ \ [\hsc]=\Psc(B)=P(B)\]
Indeed, for any $\p\in\hilb$, $[\hac]\p=\pac$ while
\beast
P(A)\p=P(A)\pac+P(A)(\psic+\ppp)=\Pac(A)\p+(\Psc(A)+\Ppp(A))\p\\
=\Pac(A)\p=\Pac(\Rr)\p=\pac
\east
The proof is similar for $[\hsc]$. The result shows that $\Pac(\D)$ and
$\Psc(\D)$ are the spectral projections $P(A\cap\D)$ and $P(B\cap\D)$,
respectively. Using Eq.(\ref{pacsc}) and the definition of $\mup$, it is
immediately seen that $\mup\ll\Pac$ if $\p\in\hac$ and $\mup\ll\Psc$ if
$\p\in\hsc$.

At last,
\[\s(H)=\overline{\spp(H)}\cup\sac(H)\cup\ssc(H)\]
where
\bea\label{spectra}
\begin{array}{lll}
\s(H)  &=&\supp P\\
\spp(H)&=&\mbox{ess.supp}\,\Ppp\\
\sac(H)&=&\supp\Pac\\
\ssc(H)&=&\supp\Psc
\end{array}\eea

It is also important to know, how to reconstruct
the spectrum from real measures. Clearly, if $\m$ is any real measure equivalent
to $P$ then in Eq.(\ref{spectra}) $P$ can be replaced by $\m$.

{\em Problem 6.} Let $\m$ be a measure and $\mup\ll\m$ for all $\p\in\hilb$.
Then $P\ll\m$.

{\em Problem 7.} $\m_{\a\p^1+\b\p^2}\ll\m_{\p^1}+\m_{\p^2}$.

{\em Problem 8.}
Let $\p^1,\p^2,...$ be a normalized
basis in $\hilb$ and $c_i>0$ for $i=1,2,...$ such that
$\sum_{i=1}^\infty c_i<\infty$. Then
\[P\sim\sum_{i=1}^\infty c_i\m_{\p^i}\]

In the particular case of $\hilb=\eltu$ and $H$ the \Sch operator, we have a
much simpler result. Let $\d^0$ and $\d^1$ be the unit vectors concentrated on
$0$ and $1$, respectively. Then
\be\label{d0d1}
P\sim\m_{\d^0}+\m_{\d^1}
\ee
Call $\m$ the measure on the right side.
$\m\ll P$ is obvious, one has to show $P\ll\m$.
Let $\D$ be such that $\m(\D)=0$. Then $\pdel\d^0=\pdel\d^1=0$ and, as a
consequence,
\[\pdel H^n\d^0=\pdel H^n\d^1=0\ \ \mbox{for $n=1,2,...$}\]
From this we can conclude that $\pdel=0$ because of the following.

{\em Problem 9.} The set of vectors $\{H^n\d^0, H^n\d^1\}_{n=0}^\infty$
is a basis in $\eltu$.\\
Clearly, Eq.(\ref{d0d1}) holds if $0$ and $1$ are replaced by any two
successive integers.

The sum of spectral measures is, in general, not a spectral measure.

{\em Problem 10.} $\m_{\p^1+\p^2}=\m_{\p^1}+\m_{\p^2}$ if and only if the two
vectors are in orthogonal $H$-invariant subspaces.

Therefore, for the \Sch operator
the right member of the relation (\ref{d0d1}) is not a spectral
measure, in general (but it {\em is} a spectral measure for the free Laplacian,
see Section 5.9).

If $H$ has a pure point spectrum, and for the basis in Problem 8
we choose an orthonormal set of eigenvectors, $\sum c_i\m_{\p^i}$
is equal to the spectral measure belonging to $\sum\a_i\p^i$, where
$|\a_i|^2=c_i$.
It is interesting to remark, that
there always exist spectral measures equivalent to $P$, even if the spectrum
is not pure point. The construction is suggested by Problem 10. Let $\subh$
be an $H$-invariant subspace. A vector $\p\in\subh$ is a {\em cyclic vector}
in $\subh$ if any $\vf\in\subh$ can be obtained as
\[\vf=\lim_{N\gotoinf}\sum_{n=1}^N\a_{Nn} (H^n\p)\]
In this case, $\m_{\vf}\ll\mup$ for all
$\vf\in\subh$, as one can see from Problems 2 and 7.
Not all $H$-invariant subspaces contain
cyclic vectors (a counterexample is a subspace belonging to a degenerate
eigenvalue), but $\hilb$ can always be written as
\be\label{decomp}
\hilb=\oplus\subh_i\ ,
\ee
a finite or
infinite direct sum of orthogonal $H$-invariant subspaces with cyclic vectors.
Let $\p^i$ be a normalized cyclic vector in the $i$th subspace and
$\a_i\neq0$,
$\sum|\a_i|^2<\infty$. Any $\vf\in\hilb$ can be written as
$\vf=\sum\b_i\vf^i$ where $\vf^i\in\subh_i$.
According to Problem 10,
\[\m_{\vf}=\m_{\S\b_i\vf^i}=\sum|\b_i|^2\m_{\vf^i}\ll\sum|\a_i|^2\m_{\p^i}
=\m_{\S\a_i\p^i}\]
and, according to Problem 6,
$P\sim\m_{\Sigma\a_i\p^i}$.

The decomposition (\ref{decomp}) is nonunique. We can start with any $\p^1$,
choose $\subh_1$ as the smallest subspace containing $H^n\p^1$
for every nonnegative
integer $n$, choose any $\p^2$ orthogonal to $\subh_1$, and so on... . The
smallest number of subspaces we need for the decomposition may be called the
{\em multiplicity} of the spectrum of $H$.
Problem 9 shows that the multiplicity of the spectrum of a
\Sch operator on $\eltu$ is $1$ or $2$.
If the multiplicity is $1$, that is,
there exists a cyclic vector (in $\hilb$),
$H$ is multiplicity free also in the sense defined in 5.5.

Multiplicity may
not be uniform on the spectrum, therefore it is appropriate to define it also
locally: Let $\D$ be a Borel set such that $P(\D)\neq0$,
and $\hilb_{\D}=P(\D)\hilb$. The decomposition
(\ref{decomp}) can be performed on $\hilb_{\D}$, and the smallest number of
subspaces we need to it is called the multiplicity of the spectrum on $\D$.
If $\D$ is an eigenvalue, we obtain the usual notion of multiplicity.
\subsection{A spectral measure for $H_0$}
Let $\hilb=\eltu$, $H=H_0=T+T^{-1}$ (recall: $T$ is the left shift). We are
going to compute explicitly $\m_{\d^0}$. The way to proceed is to find a
measure $\m$ such that
\[(\d^0, H^n\d^0)=\int E^n\dd\m(E)\ \ \mbox{for all $n$}\]
Then we can identify $\m$ with $\m_{\d^0}$.

For any unitary operator $U$
\[(\d^0, H^n\d^0)=(U\d^0, UH^n\d^0)=(U\d^0, (UHU^{-1})^nU\d^0)\]
Choose for $U$ the Fourier transformation which maps $\eltu$
onto $L_{\mathrm{per.}}^2[0,1]$: For any $\p\in\eltu$
\[(U\p)(x)=\suminf\p_ne^{2\pi inx}\ \ ,\ \ x\in[0,1]\]
In particular, $(U\d^0)(x)\equiv 1$ and $UT^{\pm1}U^{-1}$ correspond
respectively to the multiplication by $\exp(\mp2\pi ix)$; therefore $UHU^{-1}$
corresponds to the multiplication by $2\cos2\pi x$. This gives
\[(\d^0, H^n\d^0)=\int_0^1(2\cos2\pi x)^n\dd x=2\int_0^\half(2\cos2\pi x)^n\dd x
=\int_{-2}^2E^n\frac{\dd E}{\pi\sqrt{4-E^2}}\]
where we applied the substitution $E=2\cos2\pi x$. From here we recognize
\[\dd\m_{\d^0}(E)=\opt \dd E/\pi\sqrt{4-E^2} &\mbox{   if $E\in(-2,2)$}\\
                     0 &\mbox{otherwise}\finopt\]
By integration,
\[\m_{\d^0}(E)=\half +\frac{1}{\pi}\arcsin(E/2)\ \ \mbox{if $-2\leq E\leq 2$}\]
and $0$ below $-2$ and $1$ above $2$. This function is manifestly absolutely
continuous, thus $H_0$ has an absolutely continuous spectrum in $[-2,2]$.
Because of the shift-invariance of $H_0$, $\m_{\d^1}=\m_{\d^0}$ and thus
$P\sim\m_{\d^0}$. This shows that $\s(H_0)=[-2,2]$ and the spectrum is
purely absolutely continuous.

Notice that
$\d^0$ is not a cyclic vector: $H^n\d^0$, $n=1,2,...$
generate only the `even' subspace,
\[{\cal E}=\{\p: \p_{-k}=\p_k\ \ \mbox{for all $k$}\}\]
This subspace is $H_0$-invariant and the orthogonal projection onto it is not
a spectral projection (cannot be written as a function of $H_0$ only). Therefore
$H_0$ is not multiplicity free. In fact, the spectrum is uniformly
of multiplicity $2$.
\subsection{$\eltu$ versus $\ell^2(\Nn)$}
Consider the matrix (\ref{Jacobi}) of the \Sch operator on $\eltu$ and replace
the matrix elements $H_{-1,0}=H_{0,-1}=1$ by zeros. What kind of effect this
tiny modification can have on the spectrum of the modified operator $H'$?
According to the`classical' Weyl theorem ([115], Vol. IV),
\[\sess(H')=\sess(H)\]
that is, apart from isolated eigenvalues, the two spectra coincide.
The matrix of $H'$ is of block-diagonal form, so $H'$ can be written as
$H'=H_l\oplus H_r$, where $H_l$ and $H_r$ are selfadjoint operators acting
on $\ell^2(-\Nn\setminus\{0\})$ and $\ell^2(\Nn)$, respectively. Thus,
\[\s(H')=\s(H_l)\cup\s(H_r)\]
For strictly ergodic potentials $\sess(H_l)=\sess(H_r)$, and, hence,
$=\sess(H)$.

The fate of spectral types under such a `finite-rank' perturbation is partly
uncertain. Point and singular continuous spectra may
change into each other (Gordon [55],
del Rio, Makarov, Simon [122]). However, the absolutely
continuous spectrum is robust,
\be\label{sac}\sac(H)=\sac(H_l)\cup\sac(H_r)\ee
and this holds, whatever be the (selfadjoint) boundary condition at $0$,
defining $H_l$ and $H_r$.
The reason is that the singular spectrum is sensitive to the boundary condition
at $0$, while the absolutely continuous spectrum can be fully characterized
by the large-time behaviour of propagating wave packets and, hence, is
insensitive to local perturbations. In fact, it is stable under more general
(trace class) perturbations (see e.g. [77], Chapter X, Theorem 4.4).

\subsection{Asymptotic behaviour of generalized eigenfunctions. Subordinacy}
We started Section 5 with a remark about the sufficiency, for Physics,
of polynomially bounded solutions. This remark is based on the following deep
theorem:\\
The spectral resolution of the identity has an
essential support $\cal E$ such that for
every $E\in\cal E$ there exists a solution $\p$ of the \Sch equation
satisfying the following condition: For any $\d>d/4$
($d$ is the space dimension), with a suitably chosen
positive constant $c=c(\d,E)$,
\be\label{gef}
|\p(x)|\leq c(1+|x|^2)^{\d} \ \ \mbox{for all $x$ in $\Bbb R$$^d$ or
$\Bbb Z$$^d$}
\ee
The most general form of this theorem (going beyond \Sch operators)
can be found in Berezanskii's book [22]. The \Sch case in the
continuum,  with precise conditions,
assertions and proof is described in Simon's survey paper [128]
(also in [29]).
A simple proof for the $\eltu$ case, due to Lacroix,
is presented in the book of Bougerol and Lacroix [25].

Clearly, the same result is valid separately for $\Ppp$, $\Pac$ and $\Psc$
and is nontrivial for $\Pac$ and $\Psc$. It, however, does not permit to
distinguish between these two cases.

It is expected that knowing the whole family of solutions for a given energy
$E$, one can decide whether $E$ is in the spectrum and
which spectral type it belongs to. The right notion to deal with this
question is {\em subordinacy}. It was introduced by Gilbert and Pearson [52]
for the \Sch differential equation on the half-line, extended by Gilbert [51]
to the problem on $\Rr$, and by Khan and Pearson [78]
to the discrete equation on $\Nn$. A discussion of it can be found in Pearson's
book [108]. Results for the
equation on $\Zz$ can be safely deduced from these works.

For a two-sided complex-valued sequence $\p$ and an integer $N$ let
\[\|\p\|_N=(\sum_{n=0}^N\mbox{$\!\!$*}\  |\p_n|^2)^{1/2}\]
The star means that for $N<0$ the summation goes over $n=0,-1,...,N$.
Fix $E$. A solution $\p$ of the \Sch equation with energy $E$ is said to be
{\em subordinate} at $+\infty$ ($-\infty$) if for every linearly independent
solution $\vf$,
\[\|\p\|_N/\|\vf\|_N\goto 0\]
as $N\goto\pinf$ ($\minf$).
Clearly, there can be at most one
subordinate solution at either side, and it suffices to check the condition
on a single $\vf$ linearly independent of $\p$. Subordinate solutions for real
energies are real.

Below, by `solution' we mean a solution of the \Sch equation for given $E$.
Consider the following sets.
\begin{itemize}
\item $M'=\Rr\setminus\{\einr$ there exist two solutions, one
which is subordinate at $\pinf$ \phantom{$M'=$}but not at $\minf$,
the other which is subordinate at $\minf$ but not at $\pinf\}$.
\item $\Mac'=\{\einr$ every solution is nonsubordinate at $\pinf$\}\\
\phantom{\mbox{$\Mac'$\ \,}}$\cup\{\einr$ every solution is nonsubordinate at
$\minf$\}
\item $\Msing'=\{\einr$ there exists a solution which is subordinate
at $\pminf$\}
\item $\Msc'=\{\einr$ there exists a solution which is subordinate at $\pminf$
but is not \phantom{$\Msc'=\{$}in $\eltu$\}
\item $\Mpp'=\{\einr$ there exists a solution which is subordinate at $\pminf$
and is in \phantom{$\Mpp'=\{$}$\eltu$\}
\end{itemize}
Then $M'$ is an essential support for $P$ and $\Mpp'=\spp(H)$.
($M'$ is the disjoint union of $\Mac'$, $\Msc'$ and $\Mpp'$ and is a subset
of $\s(H)$.)
Furthermore, $\Pac\neq0$ if and only if $m(\Mac')>0$, in which case $\Mac'$
is an essential support for $\Pac$ (and so is
$\Mac'\cap\sac(H)$). If $\Psc\neq0$, $\Msc'$ is an
essential support for $\Psc$ (and so is $\Msc'\cap\ssc(H)$). If $J$ is a real
interval, $\Pac(J)=\Ppp(J)=0$ and $\Msc'\cap J$ is an uncountable set, then
$\Psc(J)>0$.

The above characterization shows that physicists are not quite right when
saying that in the singular continuous spectrum `all the solutions are
critical'. In fact, there is one solution (the subordinate one) which is less
(or more) `critical' than the others and shows a clear analogy with an
eigenvector. Notice, however, that a subordinate solution may not decay to
zero at $\pm\infty$.

The union giving $\Mac'$ corresponds to the union (\ref{sac}). On the
intersection of the two parts of $\Mac'$
the multiplicity of the spectrum is $2$, otherwise it is $1$ (Kac,
[75], [76]).

Recently, Al-Naggar and Pearson [3] developed further the characterization
of the absolutely continuous spectrum in the case of the differential equation
on the half-line. Due to Eq.(\ref{sac}), their result applies to the
$\eltu$ case in the form presented below.

For fixed {\em real} $E$ a {\em complex} solution $\p$ of the \Sch equation
with energy $E$ is called {\em rotating} at $\pinf$ ($\minf$) if
\[\sum_{n=0}^N\mbox{$\!\!$*}\ \p_n^2/\sum_{n=0}^N
\mbox{$\!\!$*}\ |\p_n|^2\goto0\]
as $N\goto\pinf$ ($\minf$). Notice that a complex solution for real $E$ is
the linear combination of two real solutions with complex coefficients. Let
\[\mbox{$\Mac''=\{\einr$ there is a solution which is rotating at least at
one of $\pminf$\}}.\]
Then, if $m(\Mac'')>0$, $\Pac\neq0$ and $\Mac''$ is {\em inside} an essential
support of $\Pac$.

Unfortunately, it is not known so far whether $\Mac''$ itself is an essential
support for $\Pac$.
\newsec{\Sch equation with strict\-ly er\-god\-ic po\-ten\-tials}
\subsection{Strict ergodicity}
The first step is to check minimality and unique ergodicity of the
sequence defining the potential. We discuss this question only briefly;
the reader may consult some other lectures of this School ([1],
[40], [95], [114])
and references therein. The minimality of a sequence (and then that of the hull)
is equivalent to the infinite repetition, with arbitrary precision and bounded
gaps, of every finite segment. In the case of sequences taking values from a
finite set, once minimality is known, unique ergodicity means that
the word frequencies exist (the defining limits converge).

The cosine potential of the Almost-Mathieu equation is a uniformly almost
periodic sequence: For any $\ve$ there exists a sequence of integers, $\{n_i\}$,
with bounded gaps such that for all $i$ and all $n$
\[|V_{n+n_i}-V_{n}|<\ve\]
(see e.g. [24]). This implies also strict ergodicity. A weaker
form of almost periodicity (in the mean- or Besicovitch sense,
see [24], which is equivalent
to the existence of an atomic Fourier transform of the sequence, with square
summable coefficients) also implies strict ergodicity. Sturmian sequences, or
more generally, sequences generated by the circle map with $A$ being the union
of a finite number of intervals of the type $[a,b)$
(cf. point (2) in Sec.2) belong to this class. The Fourier coefficients of
the $1$-periodic function $X_A(x)$ are easy to compute.
It is worth noticing that in this case the Fourier series converges everywhere
but may not represent the sequence in a finite number of points.
If this occurs, it is the function
$X_A$, and not the series, which defines a strictly ergodic sequence.

Some substitutional sequences, which are not Sturmian, also admit an atomic
Fourier transform. Examples are the regular paper-folding and the
period-doubling sequences. As one-sided sequences, these are strictly ergodic.

In general, for every primitive
substitution $\x$ one can build up two-sided minimal
sequences, which are also uniquely (and, hence, strictly) ergodic
([56], [102]; see also [39]). The construction goes by
concatenating a left- and a right-sided fixed point of some power of $\x$.
More precisely, one can find two letters $a$ and $b$ and an $n\geq1$ such that
$\x^n(a)=...a$, $\x^n(b)=b...$ and, with $\eta=\x^n$, both
$u=\eta^{\infty}(a)=...a$ and $v=\eta^{\infty}(b)=b...$ contain the word $ab$
and the two-sided sequence $uv$ is minimal and uniquely ergodic.
Among others, this holds for the Thue-Morse
and the Rudin-Shapiro sequences, although their Fourier transform is not atomic.
Different constructions of a two-sided sequence (for example, via symmetric
extension)
may violate minimality (the starting sequence may not be recurrent). If both
sides are related to the same substitution, the essential spectrum will not
suffer, but the singular spectral measures can be seriously perturbed.
(Compare with Section 5.10.)
\subsection{The spectrum of $H(\o)=H_0+V(\o)$}
Let us start with three remarks.

1. The shift $T$ is defined on $\eltu$ by $(T\p)_n=\p_{n+1}$. This is a unitary
operator, therefore $T^nHT^{-n}$ is unitary equivalent to
and, thus, has the same spectrum as $H$ for any finite
$n$. This obviously holds for any potential. Notice that $TH_0T^{-1}=H_0$ and
$TV(\o)T^{-1}=V(T\o)$, thus
\[T^nH(\o)T^{-n}=H(T^n\o)\]

2. If $V^k$ is a sequence of potentials converging
pointwise to a bounded
potential $V$, then $H^k=H_0+V^k$ tends strongly to $H=H_0+V$, i.e.,
for any fixed $\p\in\eltu$, $H^k\p\goto H\p$. Indeed,
\[\|(H^k-H)\p\|^2=\sum_n|V^k_n-V_n|^2|\p_n|^2\goto0\ \
\mbox{as $k\gotoinf$}\]

3. If the bounded selfadjoint operators
$H^k$ strongly converge to the bounded selfadjoint
$H$ and $\D$ is an open interval
not intersecting $\s(H^k)$ for any $k$, the spectral projections
$\hid(H^k)$ also strongly converge to $\hid(H)$ ([77],
Theorem VIII.1.15); on the other hand, they all
vanish and, hence, $\hid(H)=0$. This, however, means that $\D$ does not
intersect $\s(H)$. In short,
\be\label{RS}
\overline{\bigcup\s(H^k)}\supset\s(H)
\ee

{\em Problem 1.} If $V(\o)$ is minimal, $\s(H(\o))$ is independent of $\o$.
{\em Hint.}
Use the above remarks to prove that
for any $\o$ and $\o'$,
$\s(H(\o))\supset\s(H(\o'))$ and $\s(H(\o'))\supset\s(H(\o))$.

The $\o$-independent spectrum is denoted by $\s(H)$.

Minimality also implies that there is no isolated point in the spectrum.
Even less is sufficient:

{\em Problem 2.} If the potential is bounded and recurrent then
$\s(H)=\sess(H)$. {\em Hint.} If $E$ is
an eigenvalue and $\p$ the corresponding eigenvector, $\vf^n=T^{-k_n}\p$ with
suitably chosen almost-periods $k_n$ is a Weyl sequence.

We emphasize that the proofs of the above problems do not use ergodicity, only
minimality or recurrence.
This is the more interesting, because the same statements
were verified for
$\r$-a.e. $\o$ by using ergodicity without minimality, see Section 5.2 for the
random case.

For one-dimensional
ergodic potentials Pastur proved that any $E$ is $\r$-almost surely not an
eigenvalue (Sec. 5.2). In fact, for this holding true, we need less than
ergodicity: If $\O$ is a shift-invariant set of potentials and
$\r$ is a shift-invariant probability measure on $\O$, which is defined on
cylinder sets,
any fixed $E$ is not in $\spp(H(\o))$ with $\r$-probability $1$.
Indeed,
\[T\hieo T^{-1}=\hie(H(T\o))\]
and
\beast
\int\tr\hieo\dd\r(\o)=\int\sum(\d^n,\hieo\d^n)\dd\r(\o)\\
=\sum\int(\d^0,\hie(H(T^{-n}\o))\d^0)\dd\r(\o)
=\suminf\int(\d^0,\hieo\d^0)\dd\r(\o)
\east
All the integrals exist, the first integral takes value in $[0,1]$
($\hieo=0$ or it projects to a one-dimensional subspace of $\eltu$)
and the last sum gives $0$ or $\infty$.
The value of the first integral
is therefore $0$, implying $\hieo=0$ for $\r$-a.e. $\o$.

The content of the above statement is that $\spp(H(\o))$ changes when $\o$
changes. On the other hand, the closure of $\spp(H(\o))$ may not change:\\
For ergodic potentials there are closed sets $\S_{\xpp}$,
$\S_{\xac}$ and $\S_{\xsc}$ such that
\beast
\overline{\spp(H(\o))}=\S_{\xpp}\\
\sac(H(\o))=\S_{\xac}\\
\ssc(H(\o))=\S_{\xsc}
\east
for $\r$-a.e. $\o$ (Kunz, Souillard [91]). This theorem is the analogue of
the assertion of Problem 1,
but is a great deal more subtle than the constancy of
the spectrum. Moreover, the result surely does not hold for {\em all} $\o$,
even if the potential is strictly ergodic:
a counterexample of [71] was mentioned in Section 5.2.
\subsection{Integrated density of states}
For finite systems the integrated density of states (IDS) as a function of $E$
counts the number of eigenvalues per unit volume below $E$. Let
$\hdir$ denote the restriction of $H(\o)$ to the interval $[-L,L]$
with Dirichlet boundary condition. The IDS for $\hdir$ is
\be\label{ids}
\ids_{\o,L}(E)=\frac{1}{2L+1}\times(\mbox{ number of eigenvalues $\leq E$ of
$\hdir$})
\ee
For uniquely ergodic potentials $\ids_{\o,L}$ has an $\o$-independent limit,
$\ids(E)$, when $L$ goes to infinity.
As a limit of monotonically increasing functions, $\ids(E)$ is
monotonically increasing and can be shown to be continuous. It is, therefore,
the distribution function of a continuous measure on $\Rr$, denoted also by
$\ids$. Furthermore, it can be shown that $\supp\ids=\s(H)$.
The use of Dirichlet boundary condition in the construction is not exclusive:
any other boundary condition yielding a Hermitian restriction of $H(\o)$
leads to the same IDS. (This, however, may not be true in higher dimensional
spaces.) Obviously, the eigenvalues have to be counted with multiplicity.

The IDS for uniquely ergodic potentials can also
be obtained by dealing directly with
the infinite system. Fix any $V(\o)$ and let $\m_{\d^n}(E)$, $n\in\Zz$,
be the distribution functions of the spectral measures for $H(\o)$,
associated with the canonical basis in $\eltu$. Their average
\[\ids_L(E)=\frac{1}{2L+1}\sum_{n=-L}^L\m_{\d^n}(E)\ ,\]
which is the distribution function of a measure ($\ids_L$),
converges to $\ids(E)$ as $L$ goes to infinity ([8]).

A part of the above results is easy to understand.
If $E$ is outside the spectrum,
$f_0(\o)=\m_{\d^0}(E)$ is a continuous function on the hull and, because of
unique ergodicity, its average along trajectories exists and is independent
of $\o$: this is $\ids(E)$. If the spectrum is a Cantor set, the spectral gaps
are dense everywhere (see Section 5.3)
and $\ids(E)$ has a unique extension from the gaps to $\Rr$ into an upper
semicontinuous function. This, of course, does not explain, why is the
spectrum a Cantor set and why is $\ids(E)$ continuous.
The IDS is constant in the
gaps with values taken from a well-defined set (see Section 6.9).

Recall that the (pointwise) convergence of the distribution functions
does not imply the convergence of the measures
on {\em every} Borel set. For example, if $H(\o)$ has a pure point spectrum,
for each $E\in\spp(H(\o))$,
\[\ids_L(\{E\})\leq 1/(2L+1)\goto0=\ids(\{E\})\]
but $\ids_L(\spp(H(\o)))=1$ for each $L$ while $\ids(\spp(H(\o)))=0$.
In general, the infinite summation, which occurs because of
the $\s$-additivity of $\ids_L$, does not commute
with the limit $L\gotoinf$.

The above remark makes less surprising the observation,
that the type of the measure $\ids$ may have nothing to do with the type of the
spectral measures. For example, in the case of
random potentials, for $\r$-a.e. $\o$ the spectrum is pure point while
$H(\o)$ generates the same, continuous, IDS. In particular,
$\ids$ is absolutely continuous
for smooth $\r$. For Bernoulli distribution and sufficiently large potential
strength $\ids$ contains a singular continuous component with a support of
positive Lebesgue measure ([28]). If the spectrum is a Cantor set of zero
Lebesgue measure, a continuous $\ids$ is necessarily singular continuous:
examples are presented in Section 7. If the IDS is singular
continuous, the {\em (differential)
density of states} can be given no meaningful definition.
\subsection{IDS and Lyapunov exponent}
In Section 3.2 we introduced the Lyapunov exponent, more precisely,
$\overline{\g}_{\pm}(E,V)$ and $\underline{\g}\,_{\pm}(E,V)$ for an arbitrary
potential $V$.
According to a theorem by F\"urstenberg and Kesten [47], in the case of
ergodic potentials there exists a function $\g(E)$ such that for every fixed
$E$, for $\r$-a.e. $\o$ the four numbers coincide to give
\[\g(E,\o)=\lim_{|n|\gotoinf}\frac{1}{|n|}\ln\|T_{1\goton}(E,\o)\|\]
and $\g(E,\o)=\g(E)$.
A detailed study of many related questions, as, for example, the uniformicity
in $E$ of the convergence can be found in Goldsheid's work [53].

Let
\bea\label{FK}
S&=&\{(E,\o):\ \mbox{$\g(E,\o)$ does not exist or $\neq\g(E)$}\}\nnu\\
S_E&=&\{\o:(E,\o)\in S\}\nnu\\
S^\o&=&\{E:(E,\o)\in S\}
\eea
According to [47], for every $E$, $\r(S_E)=0$. By the Fubini theorem,
$(m\times\r)(S)=0$ and, for $\r$-a.e. $\o$, $m(S^\o)=0$.

In the theorem of F\"urstenberg and Kesten the restriction
to $\r$-a.e. $\o$ is essential, even if the potential is strictly
ergodic. A counterexample is provided by the Almost-Mathieu equation
for $\l>2$ and $\a$ a Liouville number, where for any $E\in\s(H)$ the set $S_E$,
defined above, is nonempty ([8]). However,
for potentials
generated by primitive substitutions, Hof [63] proved that $S$ is empty.

There is a remarkable formula connecting the Lyapunov exponent to the IDS,
found by Herbert and Jones [58] and Thouless [137]:
\be\label{HJT}
\g(E)=\int\ln |E-E'|\dd\ids(E')
\ee
The main observation leading to this formula is the following. For fixed $\o$
let $H^{(D)}_{[1,L]}$ denote the restriction of $H(\o)$ to the interval $[1,L]$
with Dirichlet boundary condition. It has an $L\times L$ matrix with $V(\o)_1,
...,V(\o)_L$ in the diagonal, $1$ everywhere above and below the diagonal
and $0$ elsewhere. Let $\p$ be the solution of $H(\o)\p=E\p$ with initial
condition $\p_0=0$, $\p_1=1$. Then
\[\det(E-H^{(D)}_{[1,L]})=\p_{L+1}\]
Indeed, both members are polynomials of $E$ of degree $L$, they have the same
roots and the same principal coefficient. Let $E_{1,L},...,E_{L,L}$ be the
roots, then
\[L^{-1}\ln |\p_{L+1}(E)|=L^{-1}\sum_{i=1}^L\ln |E-E_{i,L}|\]
When $L\gotoinf$,
the two sides go to the respective sides of Eq.(\ref{HJT}), apart from a set
of $(E,\o)$ of zero $m\times\r$ measure. Equation (\ref{HJT}) holds, in fact,
for all $E\in\Cc$. The proof is based on the
subharmonicity of $\g(E)$ and is due to Herman [59] and
Craig and Simon [35].
\subsection{Results on the set $\{E: \g(E)=0\}$}
Let us recall that for periodic potentials
\[\s(H)=\sac(H)=G_0\ \mbox{where $G_0=\{E: \g(E)=0\}$}\]
For $H_0$, the pure kinetic energy operator, we still have $\s(H_0)=[-2,2]$.

For almost periodic potentials Deift and Simon [38] found
\[m(G_0)\leq 4\]
with equality if and only if the potential is constant. In fact, this result
extends to any ergodic potential (Last [92]).

Ishii [68] and Pastur [106] obtained the following result:\\
Let $V(\o)$ be a bounded ergodic potential and $P^\o$ denote the spectral
resolution of the identity for $H(\o)$. Suppose that $m(G_0)>0$.
Then for $\r$-a.e. $\o$, $\Pac^\o$ is concentrated on $G_0$.\\
The proof is easy: Choose a typical $\o$, such that
$m(S^\o)=0$ (cf. Eq.(\ref{FK})). The complement of $G_0$ can be written as
$G_0^c=A\cup B$, where
\[A=\{E:\g(E,\o)=\g(E)>0\}\ \ ,\ \ B=G_0^c\cap S^\o\]
Now $\Pac^\o(B)=0$ because $\Pac^\o$ is absolutely continuous and $B$ is of
zero Lebesgue measure. On the other hand, for every
$E\in A$, either there is no polynomially bounded
solution or there is an exponentially localized solution. Therefore
$\Pac^\o(A)=0$ and, hence, $\Pac^\o(G_0^c)=0$.

There is a different way to formulate this result.
We can drop out of $G_0$ any set of zero Lebesgue measure and close the rest:
$\Pac^\o$ is still concentrated on this set which is now closed and,
therefore, contains $\sac(H(\o))$. A particular closed set to which this remark
applies is the essential closure of $G_0$.

The {\em essential closure} of a set $A$ is
\[\overline{A}\,^\xess=\{E: m(A\cap(E-\ve,E+\ve))>0\
\mbox{for all $\ve>0$}\}\]
A short inspection may convince the reader that $\overline{A}\,^\xess$
is indeed a closed set, it is inside the closure of $A$, and what it does not
contain from $A$ is a set of zero Lebesgue measure:
\[m(A\setminus\overline{A}\,^\xess)=0\]

Thus, the Ishii-Pastur theorem can be brought into the form
\[\sac(H(\o))\subset\overline{G}\,^\xess_0\]
holding for $\r$-a.e. $\o$.

What do we drop from $G_0$ when we take its essential closure?
Imagine that, apart from an absolutely continuous spectrum, $H$ has also a
singular continuous spectrum on a disjoint set $D$ of zero Lebesgue measure.
If $\g(E)=0$ on $D$, $D$ belongs to $G_0$ but not to its essential
closure. Similar is valid for a point spectrum on which the eigenvectors are
not exponentially localized.

Most remarkably, the converse of the Ishii-Pastur theorem is
also true and we have the following. \\
For $\r$-a.e. $\o$
\be\label{Koac}
\sac(H(\o))=\overline{G}\,^\xess_0
\ee
Moreover, $G_0$ is an essential support for $\Pac^\o$.

This theorem was proven by Kotani for the differential equation
([86], [87]) and
adapted by Simon to the difference equation  ([129]).

{\em Remarks.} \\
(i) $\sac(H(\o))$ is the support of $\Pac^\o$, see Eq. (\ref{spectra}),
but it may not be an essential support. In fact,
what we add to $G_0$ when taking
its essential closure, may be a set of positive Lebesgue measure, see the
example of Problem 5.6.3. \\
(ii) Since $\sac(H(\o))\cap G_0$ is also an essential support for $\Pac^\o$,
we have, in particular,
\[m(\sac(H(\o))\cap G_0)=m(G_0)\]
for $\r$-a.e. $\o$.\\
(iii) The main content of the above theorems is that the absolutely
continuous spectrum of $H(\o)$ is $\r$-almost surely nonempty if and only if
$m(G_0)>0$. The `if' is clear from the preceding remark; the `only if' holds
because $m(G_0)=0$ implies that the essential closure of $G_0$ is empty.
\subsection{The role of periodic approximants}
Given a bounded aperiodic potential $V$, one can define a sequence of periodic
approximants, $V^k$, in such a way that
$V^k$ converges to $V$ pointwise. According to the second remark made in
Section 6.2, $H^k=H_0+V^k$ then tends strongly to $H=H_0+V$ and Eq.(\ref{RS})
holds for the spectra. Our aim is to minimize the covering set on the left
side of this equation, by choosing the best periodic approximants.

The best periodic approximants of potentials of the type
$V(\o)_n=g(n\a+\o)$, where $g$ is a period-$1$ function and $\a$ is irrational,
are obtained by replacing $\a$ with its best rational approximants, $\a_k$.
For example, if $\a=(\sqrt{5}-1)/2$, $\a_k=F_{k-1}/F_k$, where $F_k$ is the
$k$th Fibonacci number.

If $V$ is a substitutional potential, $V^k$ can be chosen to be the periodic
repetition of the sequence $f(\o_n)$, evaluated on the word
$\eta^{k}(a)\eta^{k}(b)$ (cf. Sections 2 and 6.1).
\subsection{Gordon-type theorems}
Minimality sometimes implies that the potential repeats itself (exactly or
with very good precision) on three neighboring intervals,
one of which starting with $1$, and this holds
for an increasing sequence of interval lengths. In such cases, it can be shown
that the \Sch equation has no solution decaying at the infinity, and thus the
spectrum is purely continuous. The first proof of this kind was given by
Gordon [54]. Suppose that
\be\label{Go}
(V_{-L+1}...V_{-1}V_0)=(V_1...V_{L-1}V_L)=(V_{L+1}...V_{2L-1}V_{2L})
\ee
Then
\[T_{-L+1\goto0}=T_{1\goto L}=T_{L+1\goto 2L}=A_L\]
and, applying the two sides of the Caley-Hamilton equation
\be\label{CH}A_L^2-(\tr A_L)A_L+I=0\ee
to the vectors $\P_{-L}$ and $\P_0$ (cf. Eq.(\ref{vector})), one finds that
for any solution $\p$ of the \Sch equation
\[\max\{\|\P_{-L}\|,\|\P_L\|,\|\P_{2L}\|\}\geq\half\|\P_0\|\]
Clearly, if (\ref{Go}) holds for an increasing sequence $L_n$, no solution
can decay on both sides, so there can be no eigenvector to $H$.
Let us see two examples.\\
(i) Avron,Simon [8]: For the Almost-Mathieu equation, for $\l>2$
and $\a$ Liouville number (for every positive integer $k$ there are integers
$p_k$ and $q_k$ such that $|\a-p_k/q_k|\leq k^{-q_k}$), for a.e. $\o$ the
spectrum is purely continuous. In this case, Eq.(\ref{Go}) does not hold exactly
but with an extremely large precision. The spectrum is, in fact, purely singular
continuous: this comes from the Ishii-Pastur-Kotani theorem, because $\g(E)>0$
for all $E$ ([4]).\\
(ii) Delyon, Petritis [43]: For circle potentials (see Section 2) with $A$
being an interval, for
every $\l$, a.e. $\o$ and a.e. $\a$ the spectrum is
purely continuous. In order that Eq.(\ref{Go}) hold true, some weak condition on
the continued fraction expansion of $\a$ has to be imposed. This limits the
result to almost every $\a$.

A variant of the Gordon theorem uses only two intervals. Suppose that in
Eq.(\ref{Go}) the second equality is verified for an increasing sequence
$L_n$. This situation arises, for example, with substitutional potentials,
if the substitutional sequence starts with a square.
Applying the two sides
of Eq.(\ref{CH}) to $\P_0$, one obtains
\[\max\{|\tr A_L|\|\P_L\|,\|\P_{2L}\|\}\geq\half\|\P_0\|\]
This yields the absence of decaying solutions at $+\infty$ for energies such
that
\[\tr T_{1\goto L_n}(E)\]
is a bounded sequence. In some cases one can
show that this set of energies is just the spectrum (see [135]
 and Section 7).
\subsection{Kotani theorem for potentials of finite range}
Any periodic potential on $\eltu$ is of finite range, i.e., takes on a finite
number of different values. Any \Sch operator on $\eltu$
with a periodic potential has a purely
absolutely continuous spectrum. Therefore,
the following theorem by Kotani [88] may be surprising. Let
$V(\o)$ be an ergodic nonperiodic finite-range
potential, $\r$ the corresponding measure on the
hull of $V$. Then, for $\r$-a.e.\ $\o$, $H(\o)$ has a purely singular spectrum.

This theorem is at the origin of many results on \Sch operators with circle-
and substitutional potentials; we discuss them in the next section.
It has long been an open question, whether the
restriction to $\r$-a.e. $\o$ can be dropped, and whether the spectrum is purely
singular continuous for every $\o$, if the potential is strictly ergodic.
A recent result by Hof, Knill and Simon [64] goes in this direction: The
authors show that for strictly ergodic potentials, either $H(\o)$ has pure point
spectrum for all $\o$ or there is an uncountable dense set, although of zero
$\r$-measure, in the hull for which the spectrum is purely singular continuous.
This latter case is shown to be realized for circle potentials if $\a$ is
irrational and $A$ is a half-open interval. This includes Sturmian potentials.
Furthermore, for potentials generated by primitive substitutions, the Lyapunov
exponent $\g(E,\o)$ is independent of $\o$ (Hof, [63]); Kotani's theorem
then implies that the spectrum is purely singular for all $\o$.
\subsection{Gap labeling}
Gap labeling is a book-keeping for spectral gaps.
The IDS naturally assigns a number to each gap: the (constant) value $\ids(E)$
for $E$ in the gap. The question is then to characterize this set of numbers.
The first example of gap labeling
for an almost periodic \Sch equation was given by Johnson and Moser [73] in
the continuum case. The method used in the discrete case is quite different.

{\em Problem 1.} Let $V$ be an $L$-periodic potential and suppose that $H=H_0+V$
has no missing gap. Show that $\ids(E)=k/L$, $k=1,2,...,L-1$ in the gaps.

For strictly ergodic
potentials, the values of the IDS in the gaps
are taken from (but do not necessarily exhaust)
a set $S$, determined by an algebraic theory, the K-theory.
The application of K-theory to gap labeling was
developed by Bellissard, Lima, Testard, Bovier and Ghez  ([19],
[16]).
A detailed discussion can be found in [13], [14].

Sometimes it is possible to describe $S$ quite
explicitly.  For example, according to Bellissard, Bovier and Ghez ([14],
[13], [16], [27]), for circle and
substitutional potentials $S$ is the module (smallest additive group)
containing the word frequencies of $V(\o)$, restricted to the interval
$[0,1)$. This module
contains the integers (the sum of the frequencies of cylinders with common
base ($n$ and $k$ in Eq.(\ref{cyl})) is $1$), therefore $S$ is
a group with respect to addition modulo $1$, like for periodic potentials.

For Sturmian potentials this gives
\[S=\{k\a+m(1-\a): k,m\in\Zz\}\cap[0,1)=\{k\a+m: k,m\in\Zz\}\cap[0,1)\]
Notice that for $\a=K/L$, where $K$ and $L$ are relatively primes, this
gives
\[S=\left\{\frac{1}{L},\frac{2}{L},\ldots,\frac{L-1}{L}\right\}\;,\]
in accordance with the assertion of the Problem above and the example (ii)
in Section 4.
If $\a$ is irrational, $S$ is countably infinite and dense in $[0,1]$.
If there exist, indeed, gaps corresponding to values in a subset $S'$ of
$S$ which is still dense in $[0,1]$, the spectrum
is necessarily a Cantor set and the IDS is continuous.

In our discussion of the periodic \Sch equation (Section 4) we introduced the
notion of a missing gap (or closure of a gap). When there are no intervals
in the spectrum,
as is often the case with strictly ergodic potentials, what one can
unambiguously assert is the {\em absence} of missing gaps or the
{\em completeness of gap labeling}.
Surely, there is no missing gap if
\be\label{ranids}
\Ran\ids\equiv\{\ids(E):\ \mbox{$E$ is in a spectral gap}\}=S\ ,
\ee
the set of all the admissible values.
(Notice that each admissible value is taken on in at most a single gap.)
As in the periodic case, there is no general method to check whether or not
all the gaps open. This is known to hold true in a few cases,
like the period doubling potential or the Fibonacci potential with $\l>4$.

{\em Example.}
For the classical Cantor set, in the $k$th gap on
the $n$th level ($n=0,1,...;\ k=1,...,2^n$)
the Cantor function $\a(x)=(2k-1)/2^{n+1}$. These
numbers form an additive group modulo $1$. The same values are taken by the
IDS of the hierarchical Hamiltonian ([90], see also [89]) in the domain
of the parameters where the potential is limit periodic. In this case the
gap labeling is complete.
\newsec{\Sch equation with Sturmian and sub\-sti\-tu\-tional po\-ten\-tials}
\subsection{Fibonacci potential}
Fibonacci substitution was the first to be used to define a two-valued
potential (see Section 2) and to study the spectral problem of the
corresponding \Sch operator (Kohmoto, Kadanoff and Tang, Ostlund et al.
[79], [80], [81], Casdagli [30], S\"ut\H{o} [135],
[136]). This is also the
first example where the trace map ([2], [84], [111]) was fully
exploited.

The Fibonacci sequence, as any minimal sequence, has almost-periods: these
are the Fibonacci numbers $F_n$ ($F_0=F_1=1, F_{n+1}=F_n+F_{n-1}$).
Choose, for example,
\be\label{Fib}V_n=\l(\lfloor(n+1)\a\rfloor-\lfloor n\a\rfloor)\ee
then the almost-periodicity is expressed by the equations
\beast
V_{l+F_n}&=&V_l\ \ \mbox{if $n\geq3$ and $1\leq l\leq F_n$}\\
V_{l-F_{2n}}&=&V_l\ \ \mbox{if $n\geq1$ and $1\leq l\leq F_{2n+1}$}
\east
The transfer matrices over the almost-periods replace the single transfer
matrix $T_{1\goto L}$ in the $L$-periodic case.
For an $L$-periodic potential
\be\label{per}T_{1\goto 2L}=T_{1\goto L}^2\ee
Let
\[M_n=T_{1\goto F_n}\]
For the Fibonacci potential Eq.(\ref{per}) is replaced by
\[M_{n+1}=M_{n-1}M_n\]
This, together
with $\det M_n=1$, implies for $\t_n=\tr M_n$ the recurrence relation
\[\t_{n+2}=\t_{n+1}\t_n-\t_{n-1}\]
The initial conditions (with the choice (\ref{Fib}))
\[\t_{-1}=2\ ,\ \t_0=E\ ,\ \t_1=E-\l\]
contain only $2$ parameters, and this indicates that the recurrence must have
a nontrivial invariant. Indeed,
\[\t_{n+1}^2+\t_n^2+\t_{n-1}^2-\t_{n+1}\t_n\t_{n-1}-4=(\t_1-\t_0)^2=\l^2\]
From the trace map and the invariant, the following results can be deduced
([135], [136]).
\begin{enumerate}
\item $\s(H)=\{E:\mbox{$\{\t_n(E)\}_{n=1}^{\infty}$ is bounded}\}$\\
      (Remember that for an $L$-periodic potential
      $\s(H)=\{E:|\tr T_{1\goto L}(E)|\leq2\}$.)
\item The spectrum of $H(\o=0)$ is purely
      continuous (the proof is of Gordon-type with two intervals).
\item $\g(E,\o=0)=0$ for all $E\in\s(H)$.
\item The Lebesgue measure $m(\s(H))=0$ (found by confronting the
      Ishii-Pastur-Kotani theorem with Kotani's theorem
      for potentials of finite range). As a consequence,
      $\s(H)$ is a Cantor set (because $\s(H)$ contains no isolated point).
\item The spectrum of $H(\o=0)$ is purely singular continuous.
\item For all $E\in\s(H)$ all solutions of $H(\o=0)\p=E\p$ are polynomially
      bounded (Iochum, Raymond, Testard [66], [67]).
\end{enumerate}
It is an open question whether the spectrum is purely singular continuous
for all $\o$.
Let us recall that $TH(\o)T^{-1}=H(T\o)=H(\o+\a)$ is unitary equivalent to
$H(\o)$, therefore the singular continuity holds
true for $\o=k\a\,$(mod $1$), $k\in\Zz$,
which is a countable dense set in $[0,1]$.
Due to [64], this result has recently been extended to an uncountable dense
set of $\o$, still of zero measure (cf. Section 6.8).
\subsection{General Sturmian potentials}
The case of Sturmian potentials for arbitrary irrational $\a$ was investigated
by Bellissard, Iochum, Scoppola and Testard [17]. The methods applied to
the Fibonacci case can be extended to treat the general problem,
even though this is technically more involved.
All the results
found for the Fibonacci potential remain valid and the same questions are
unanswered.

There is an interesting numerical work by Ostlund and Kim [OK] and a rigorous
study by Bellissard, Iochum and Testard [18] on the $\a$-dependence of the
spectrum of $H=H_0+V^{(\a)}$, where
\[V^{(\a)}_n=X_{[0,\a)}(n\a)\ ,\]
cf. Section 2. For rational
$\a$ the potential is periodic and the spectrum is the union of a finite number
of intervals. When approaching a rational $\a$ from above and from below, the
numerical plot of the energies belonging to the spectrum clearly reveals a
discontinuity, see Figure 2. [18] show that this reflects the discontinuity
of the characteristic function which generates the potential. The gap edges
vary continuously on irrational $\a$'s. For a rational value $r$,
\[\lim_{\a\uparrow r}V^{(\a)}\neq\lim_{\a\downarrow r}V^{(\a)}\]
and none of them equals $V^{(r)}$. In fact, the two limits are not periodic,
only ultimately periodic (as if we took a periodic sequence, cut off
a finite segment which is not a period and glue the two infinite pieces
together). Both yield the same essential spectrum as $H_0+V^{(r)}$
and both create (different) isolated eigenvalues in the gaps.
\subsection{Period doubling potential}
The potential is generated by the period doubling substitution $\x(a)=ab$,
$\x(b)=aa$. The corresponding \Sch equation was studied by Bellissard,
Bovier and Ghez [15], [27].
The trace map is a fundamental tool of the
analysis. Results 1.-5. valid for the Fibonacci potential hold true:
The spectrum is a Cantor set of zero Lebesgue
measure, it is purely singular continuous (again, not known for all elements
of the hull) and the Lyapunov exponent vanishes
in the spectrum. There is a rather detailed knowledge about the gaps, their
behavior as a function of the potential strength.
Gap labeling is complete: all admissible values are taken by the IDS in the
gaps.
\subsection{Thue-Morse potential}
Historically, the first studies were done by Axel et al.\ on the phonon
frequency spectrum of the harmonic chain with masses
generated by the Thue-Morse substitution $\x(a)=ab$, $\x(b)=ba$.
In [9], [10], [11]
we find the proof
that the phonon spectrum is a Cantor set,
a numerical work on its box dimension suggesting zero
Lebesgue measure, gap labeling  and a study of
generalized eigenfunctions. This latter established the existence of extended
states for a dense set in the phonon spectrum.

The \Sch equation with
the potential generated by this substitution has also been
widely studied, namely by Bellissard [12],
Delyon and Peyri\`ere [43] and Bovier and Ghez [26],
[27]. Valuable  numerical work was done by Riklund et al.~[121].
Bellissard [12] identified the
set $\Ran\ids$ (cf. (\ref{ranids})) and studied the dependence of the gap widths
on the potential strength. Delyon and Peyri\`ere [43] proved the absence of
decaying solutions and, hence, the continuity of the spectrum. They showed that
the generalized eigenfunctions are not `too small' on a geometric progression.
The continuity  of the spectrum was obtained as a byproduct also in
[15], [27].
Bovier and Ghez [26],
[27] proved,  in  a more general context (see next section),
that the spectrum is of zero Lebesgue measure. With the continuity  this implies
that the spectrum is purely singular continuous.
\subsection{Systematic study of substitutional potentials}
Bovier and Ghez [26],
[27] succeeded to find a  general condition on primitive
substitutions, assuring that the corresponding \Sch operator has a spectrum of
zero Lebesgue measure. Their work deeply exploits the Kol\'ar-Nori results on
trace maps ([84], see also [111]).

The action of a substitution $\x$ can be defined
on transfer matrices and on traces of transfer matrices. $\x$ acting on
traces is called a {\em trace map}. $\x$ carries a letter $a$ of the alphabet
$\cal A$
into a word, the transfer matrix corresponding to $a$
into a product, i.e. a {\em monomial}, of transfer matrices
in reversed order, and the trace of the transfer matrix
into a {\em polynomial} of traces of transfer matrices belonging to letters
of an enlarged alphabet $\cal B$. These generalized letters are special
words over the original alphabet $\cal A$, not containing repeated letters of
$\cal A$. Therefore $\cal B$ is also finite, and
one can see that the trace map is closed in the following
sense:\\
$\x$ carries the trace belonging to any letter of $\cal B$ into a
polynomial of traces belonging to letters of $\cal B$.

%
%
Let $b_1,...,b_q$ be the letters of $\cal B$ and $x_i$ a real variable
associated with $b_i$.
The idea of Bovier and Ghez is to retain
the highest degree monomial
\[\tilde{f}_i(x_1,...x_q)=\prod_{j=1}^q x_j^{k_{ij}}\]
of the polynomial corresponding to $b_i$ and to define
a substitution $\phi$
on $\cal B$ such that $\phi(b_i)$ contains $b_j$ $k_{ij}$ times.
Based on earlier experience, namely with the Fibonacci substitution,
one may expect that, by imposing
some conditions on $\phi$, it is possible to
control the high iterates of the trace map
and, hopefully, the spectrum. The right notion is {\em semi-primitivity},
a property which is somewhat weaker than primitivity (see
[26] for the definition) and is
straightforward to verify. The authors prove that, if $\x$ is primitive,
$\phi$ is semi-primitive and the substitutional sequence contains the word $bb$
for some $b\in\cal B$, $\s(H)$ is a set of zero Lebesgue measure. Although the
proof is more complicated than in the cases discussed earlier, the main idea is
again to show that the Lyapunov exponent vanishes in the spectrum (the existence
of the word $bb$ is used here) and, then, to confront the Ishii-Pastur-Kotani
theorem with the Kotani theorem for finite-ranged potentials. If the
substitutional sequence starts with a square, the two-interval version of the
Gordon theorem yields also that the spectrum is purely singular continuous.

The conditions of the theorem are fulfilled by many named substitutional
sequences as, for example, the
Fibonacci, Thue-Morse, period doubling, binary non-Pisot, ternary non-Pisot
and circle sequences. Apart from the binary non-Pisot sequence, singular
continuity is also verified in the cases listed above (but it is not proved
for the other potentials in the hull).
A notable exception is the Rudin-Shapiro substitution which is primitive and
therefore the spectrum is purely singular ([64]), but
for which the substitution $\phi$ is not semi-primitive and, hence, it is not
known whether $\s(H)$ is of zero Lebesgue measure.
\section{Solutions to the problems}

{\bf 2.1}

The proof is based on the observation that for bounded sequences pointwise
convergence is equivalent to convergence in the metric $d$.

(i) 1. implies 2.

Indeed, $s\in\O(s)=\O(t)$.

(ii) 2. implies 3.

$\O(s)$ is bounded: for all $t\in\O(s)$,
$d(s,t)\leq 6\sup|s_n|$.

Suppose that $s$ is not almost-periodic. Then
there exists an $\ve>0$ and an infinite sequence of real intervals
$(a_k,b_k)$ such that $b_k-a_k\gotoinf$ and $d(T^n\/s,s)>\ve$ for every
integer $n\in\cup(a_k,b_k)$. Choose an integer $n_k$ in $(a_k,b_k)$ such
that $n_k-a_k\gotoinf$ and $b_k-n_k\gotoinf$. Since $\O(s)$ is closed
and bounded (compact in the topology generated by $d$), $T^{n_k}\/s$ has at
least one limit point $t$ in $\O(s)$. Fix an arbitrary integer $n$.
We have the triangle inequality
\[d(s,T^n\/t)\geq d(s,T^{n_k+n}\/s)-d(T^{n_k+n}\/s,T^n\/t)\;.\]
For $k$ sufficiently large, $n_k+n\in(a_k,b_k)$ and thus
$d(s,T^{n_k+n}\/s)>\ve$. On the other hand,
$d(T^{n_k}\/s,t)\goto0$ implies
that $d(T^{n_k+n}\/s,T^n\/t)\goto0$ as $k\gotoinf$. It follows that
$d(s,T^n\/t)\geq\ve$ for all integers $n$. Hence, $s$ cannot be in $\O(t)$,
which contradicts 2.

(iii) 3. implies 2.

Let $t\in\O(s)$, $t=\lim_{j\gotoinf}T^{i_j}\/s$,
where $\lim$ means pointwise limit. Fix an $\ve>0$ and let $\{n_k\}$ be
the almost-periods with gaps smaller than $\ell_\ve$. Given $j$, choose
$k$ such that $n_{k-1}<i_j<n_k$. Then $m_j\equiv n_k-i_j<\ell_\ve$ and
\beast
d(s,T^{m_j}\/t)&\leq& d(s,T^{n_k}\/s)+d(T^{n_k}\/s,T^{m_j}\/t)\\
             &\leq& \ve + d(T^{m_j}T^{i_j}\/s,T^{m_j}\/t)
\east
Take the limit $j\gotoinf$. Since $0<m_j<\ell_\ve$, there will be an
$m\in(0,\ell_\ve)$ which occurs infinitely many times among the $m_j$'s.
Taking the limit only on the subsequence of $j$ values for which $m_j=m$,
we obtain $d(s,T^m\/t)\leq\ve$. Since $\ve$ was arbitrary, this proves that
$s\in\O(t)$.

(iv) 2. implies 1.

Suppose $t\in\O(s)$. Then $T^n\/t\in\O(s)$ for every
integer $n$, because $\O(s)$ is shift-invariant. $\O(s)$ being closed,
this implies $\O(t)\subset\O(s)$. On the other hand, according to 2.,
$s\in\O(t)$. Interchanging $s$ and $t$, the above argument yields
$\O(s)\subset\O(t)$.

{\bf 2.2}

Let $A\subset\O(s)$
be a $T$-invariant set. One has to show that $\r(A)=0$ or $1$.
Suppose the opposite, i.e., $0<\r(A)<1$. Define a probability measure $\m$
by setting
\[\m(B)=\r(A\cap B)/\r(A)\]
It is easy to verify that $\m$ is a $T$-invariant probability;
on the other hand, it
differs from $\r$ (e.g., on $\O(s)\setminus A$).
This contradicts the uniqueness of $\r$.

{\bf 2.3}

Let $\o\in\O$. If $\{T^n\o\}_{n=0}^\infty$ is an infinite set, the sequence
contains no repetition. Therefore
\[1\geq\r(\{T^n\o\}_{n=0}^\infty)=\sum_{n=0}^\infty\r(\{T^n\o\})
=\sum_{n=0}^\infty\r(\{\o\})\]
and hence $\r(\{\o\})=0$.

{\bf 3.2.1}

The proof goes by induction. $\tr A^0=\tr I=2$.
Multiplying the Caley-Hamilton equation
\[A^2-(\tr A)A+I=0\]
with $A^{k-2}$ and taking the trace one finds
\[\tr A^k=\tr A\,\tr A^{k-1}-\tr A^{k-2}\]
in which one may recognize the recurrence relation for the Chebyshev
polynomials. Setting $\tr A=2\cos\a$, the solution of the recurrence is
\[\tr A^k=2\cos k\a\].

{\bf 5.1.1}

Let $H\p=E\p$ where $\p$ is subexponential: for any $a>0$ there exists a $b>0$
such that
\[|\p_k|<be^{a|k|}\ \ \mbox{for all $k$,}\]
but $\p$ is not in $\eltu$. It follows that either
$\sum_{k=0}^\infty|\p_k|^2=\infty$ or $\sum_{k=-\infty}^0|\p_k|^2=\infty$.
Suppose, for instance, the first.
Define
$\p^n=\p/(\sum_{l=0}^n|\p_l|^2)^{1/2}$ for $n>0$. Then
$H\p^n=E\p^n$, and $\vf^n$, given by
\[\vf^n_k=\opt\p^n_k&\mbox{if $0\leq k\leq n$}\\
               0&\mbox{otherwise}\finopt\]
is a normalized vector for all $n$. We show that a suitable subsequence
of $\{\vf^n\}_{n=1}^\infty$
is a Weyl sequence. It is clear that for all $k$, $\vf^n_k\goto 0$ with
increasing $n$, so $\vf^n$ goes to zero weakly. On the other hand,
\[(H\vf^n)_k=E\vf^n_k\ \ \mbox{if $k\neq -1,0,n,n+1$}\]
This implies that
\[\ve^n_k=0\ \ \mbox{if $k\neq -1,0,n,n+1$}\]
(cf. Eq.(\ref{Weyl})). It is easy to check that
\[\ve^n_{-1}=\p^n_0\ \ \ \ve^n_0=-\p^n_{-1}\ \ \ \ve^n_n=-\p^n_{n+1}\ \ \
\ve^n_{n+1}=\p^n_n\]
Therefore
\[\|\ve^n\|^2_{\ell^2}=\|\P^n_{-1}\|^2+\|\P^n_n\|^2\]
where the vectors on the right correspond to the definition (\ref{vector}).
The first term goes to zero as $n$ increases, the second, in general, not.
However, there exists a subsequence $n_k$ such that
$\|\P^{n_k}_{n_k}\|\goto0$ with increasing $k$. Indeed, if
the opposite were true, one could find a positive constant $c$ such that
$\|\P^n_n\|^2>c$ for all $n>0$. Now
\[c<\|\P^n_n\|^2=\|\P_n\|^2/\sum_{k=0}^n|\p_k|^2\leq2\|\P_n\|^2/\sum_{k=0}^{n-1}
\|\P_k\|^2\]
and therefore
\[\|\P_n\|^2>\cfel\sum_{k=0}^{n-1}\|\P_k\|^2>\cfel(\cfel+1)\sum_{k=0}^{n-2}
\|\P_k\|^2>\cdots>\cfel(\cfel+1)^{n-1}\|\P_0\|^2\]
which contradicts the supposed subexponential nature of $\p$.
The Weyl sequence we were looking for is $\{\vf^{n_k}\}_{k=1}^\infty$.

{\bf 5.2.1}

In general, let $E$ be any accumulation point of the eigenvalues of $H$, say,
$E_n\goto E$ as $n\gotoinf$. Let $\p^n$ be the corresponding orthonormal
eigenvectors. Then $\p^n$ form a Weyl sequence. First, they go to zero weakly:
for any vector $\vf\in\hilb$, the Bessel inequality
\[\|\vf\|^2\geq\sum_{n=1}^\infty|(\p^n,\vf)|^2\]
implies that $(\p^n,\vf)\goto0$ with increasing $n$. Second,
\[H\p^n=E_n\p^n=E\p^n+(E_n-E)\p^n\]
and $\|(E_n-E)\p^n\|=|E_n-E|\goto0$.

{\bf 5.2.2}

Choose a sequence $n_k$ such that $V_{n_k}\goto E$ as $k\gotoinf$. If $\d^n$
is the unit vector concentrated on $n$,
\[\|(E-V)^{-1}\d^{n_k}\|=1/|E-V_{n_k}|\gotoinf\ ,\]
so the inverse of $E-V$ is unbounded. By definition, this means that $E$ is
in the spectrum.

One can also construct a vector $\p$ such that the solution of $(E-V)\vf=\p$
for $\vf$, given by
\[\vf_n=\p_n/(E-V_n)\ ,\]
is not square-summable.
Write $E=\cos2\pi\theta$, then trivially
\[|E-V_{n}|\leq 2\pi\min_{p\in\Zz}|n\a-\theta-p|\]
On the other hand, by the Kronecker theorem (Theorem 440 in
[57]),
there exists an increasing sequence $n_k$ of positive integers and a sequence
$p_k$ of integers such that
\[|n_k\a-\theta-p_k|<3/n_k\]
The example is obtained by
choosing $\p_n=E-V_n$ for $n=n_k$ and zero otherwise: $\|\p\|^2<6\pi^4$ but
$\vf_{n_k}=1$ for all $k$, so $\vf$ is not square-summable.

{\bf 5.6.1}

If $x$ is not in $\suppm$ then for sufficiently small $\ve>0$,
$\m(x+\ve)-\m(x-\ve)=\m((x-\ve,x+\ve))=0$ and, hence, $x$ is not a point of
increase of the distribution function. If $x\in\suppm$ then for every $\ve>0$,
\[\m(x+\ve)-\m(x-\ve)=\m((x-\ve,x+\ve])\geq\m((x-\ve,x+\ve))>0\ ,\]
that is, $x$ is a point of increase.

{\bf 5.6.2}

Let $\m,\n$ be {\em ac} measures, $A$ a common essential support.
Suppose that $\m(B)=0$.
$\n(B\cap A^c)=0$ because $A$ is an essential support of $\n$.
$m(B\cap A)=0$, because $B\cap A\subset A$, an essential support of $\m$,
and $\m(B\cap A)=0$. But $\n$ is {\em ac}, so $\n(B\cap A)=0$. We found
$\n\ll\m$. The opposite is obtained by interchanging $\m$ and $\n$.

{\bf 5.6.3}

(i) Let $C$ be a `thick' Cantor set (i.e., $m(C)>0$),
\[C^c=\cup_{n=0}^\infty C_n^c=\cup_{n=0}^\infty\cup_{k=1}^{k_n}J_{nk}\]
where $J_{nk}$ are the open
intervals appearing in the construction of $C$ (cf. Sec.5.3).
Let $f_{nk}(x)$ be a continuous function which is strictly positive on $J_{nk}$
and vanishes outside $J_{nk}$, and
\[\sum_{n=0}^\infty\sum_{k=1}^{k_n}\int f_{nk}(x)\dd x<\infty\]
Then $\sum_n\sum_k f_{nk}$ is the density of an absolutely continuous measure
$\m$. $\suppm=\Rr$ (because the closure of $C^c$ is $\Rr$), and
an essential support of $\m$ is $C^c$. So $\m(C)=0$ but $m(C)>0$, therefore
$\suppm$ is not an essential support of $\m$.

(ii) For $0<p<q$, $p$ and $q$ integers which are relatively primes, let
$\m(\{p/q\})=2^{-p-q}$. $\suppm=[0,1]$ while the rational numbers in $(0,1)$
form an essential support. $\m([0,1]\setminus\Qq)=0$ but
$m([0,1]\setminus\Qq)=1$, so the support is not an essential support.

(iii) Let $\m$ be concentrated on a discrete set (whose only accumulation
points can be $\pminf$), for instance,
$\m(\Rr)=\sum_{n=1}^\infty\m(\{n\})$. Then $\suppm$ is the smallest essential
support.

{\bf 5.6.4}

(i) If $\m(x)$ is absolutely continuous ({\em ac}) on any finite interval
then the measure $\m$ is {\em ac}.\\
Indeed,
let $A$ be a Borel set, $m(A)=0$ and suppose first that $A$ is covered by a
finite open interval $J$.
Fix $\ve>0$ and let $\d>0$ correspond to $\ve$ (cf. Eq.(\ref{abscont})).
According to the definition of a set of zero Lebesgue measure,
one can find an open set $O$ such that $A\subset O$ and $m(O)<\d$. $O$ is the
union of disjoint open
intervals $(a_1,b_1),(a_2,b_2),...$, all in $J$,
so that $A\subset\cup (a_i,b_i)$ and $\sum (b_i-a_i)<\d$. Thus,
\[\m(A)<\sum\m((a_i,b_i))=\sum(\m(b_i)-\m(a_i))<\ve\]
Since $\ve$ is arbitrary, $\m(A)=0$.

If $A$ is not covered by a finite interval,
it is covered by countably many finite intervals $J_k$. $\m(A\cap J_k)=0$ for
all $k$, thus $\m(A)=0$.

(ii) If $\m$ is an {\em ac} measure, i.e., $\m\ll m$, then $\m(x)$ is an
{\em ac} function on any finite interval.\\
We show this by proving that, given a finite interval $J$,
for any $\ve>0$ there exists a $\d>0$ such that for Borel sets $A\subset J$,
$m(A)<\d$ implies $\m(A)<\ve$. Suppose this does not hold true. Then there
exists an $\ve>0$ such that for all $n\geq1$ and for suitably chosen
$A_n\subset J$, $m(A_n)<1/2^n$ but $\m(A_n)\geq\ve$. Let
$B_n=\cup_{k=n+1}^\infty A_k$. Now $m(B_n)\leq1/2^n$ and
$\m(B_n)\geq\m(A_{n+1})\geq\ve$. Since $B_n\subset J$ for all $n$ and $B_n$
is a decreasing sequence, it has a limit $B\subset J$ for which $m(B)=0$ and
$\m(B)\geq\e$. This contradicts $\m\ll m$.

{\bf 5.6.5}

Suppose $\m$ is a singular measure. Then there is some set $A$ of zero Lebesgue
measure such that $\m$ is concentrated on $A$. The intersection of any
essential support with $A$ yields
an essential support of zero Lebesgue measure. If $B$ is an essential support
and $m(B)>0$ then $\m(B\setminus A)=0$ and $m(B\setminus A)>0$, a contradiction.

{\bf 5.6.6}

The definition of $\mpp$ is constructive, see Eq.(\ref{mpp}), therefore the
{\em pp} part is unique. Let
\[\m=\mpp+\msc^1+\mac^1=\mpp+\msc^2+\mac^2\]
If $\msc^1\neq\msc^2$, there exists a set $\D$ such that $m(\D)=0$ and
$\msc^1(\D)\neq\msc^2(\D)$. But then $\mac^1(\D)\neq\mac^2(\D)$ which
contradicts $\mac^1(\D)=\mac^2(\D)=0$.

{\bf 5.7.1}

Let $C$ be the middle-thirds Cantor set, $\a$ the {\em sc} measure whose
distribution function is the Cantor function. Let $\m$ be a {\em pp} measure
whose smallest essential support is the set of boundary points $\{c_{ni}\}$
of the complement of $C$, cf. Sec. 5.3. Then $\supp\m=\supp\a=C$ and $C$ is
also an essential support for both measures.

{\bf 5.8.1}

The following statements are equivalent:
$\mup$ is concentrated on $A$, $\mup(\Rr)=\mup(A)$, $\|\p\|^2=\|P(A)\p\|^2$,
$\|(I-P(A))\p\|=0$, $(I-P(A))\p=0$, $P(A)\p=\p$.

If $\p\in P(A)\hilb$ then
$P(A)\p=\p$, so $\mup(A^c)=0$ and $\supp\mup\subset\overline{A}$.

{\bf 5.8.2}

Let $A$ be any Borel set such that $\mup$ is concentrated on $A$.
Then from Problem 5.8.1, $P(A)\p=\p$.
$H$ commutes with $P(A)$, therefore $P(A)H\p=H\p$. From Problem 5.8.1 it
follows that $\m_{H\p}$ is concentrated on $A$. By definition, we get
$\m_{H\p}\ll\mup$.

If $\mup$ is singular, it is concentrated on a set $A$ such that $m(A)=0$.
Therefore $\m_{H\p}$ is also singular. If $\mup$ is continuous,
for any $E\in\Rr$, $\mup(\{E\})=\|\pe\p\|^2=0$, therefore $\pe\p=0$. Hence,
\[\m_{H\p}(\{E\})=\|\pe H\p\|^2=\|H\pe\p\|^2=0\ ,\]
that is, $\m_{H\p}$ is
continuous. So $\m_{H\p}$ is {\em sc} if $\mup$ is {\em sc}.
If $\mup$ is {\em ac} then $\m_{H\p}$ is {\em ac} because $m(A)=0$ implies
$\mup(A)=0$, which implies $\m_{H\p}(A)=0$.

{\bf 5.8.3}

Let $A,B,C$ be disjoint essential supports respectively for $(\mup)_\xpp,
(\mup)_\xac,(\mup)_\xsc$. $\mup$ is concentrated on $A\cup B\cup C$. From
Problem 5.8.1,
\[\p=P(A\cup B\cup C)\p=P(A)\p+P(B)\p+P(C)\p\equiv\ppp+\pac+\psic\]
Now
\[P(A)P(B)=P(A)P(C)=P(B)P(C)=0\]
therefore $\ppp,\pac,\psic$ are pairwise orthogonal,
and for any Borel set $\D$
\beast
\mup(\D)=\|P(\D)\p\|^2
&=&\|P(\D)P(A)\p\|^2+\|P(\D)P(B)\p\|^2+\|P(\D)P(C)\p\|^2\\
&=&\m_\ppp(\D)+\m_\pac(\D)+\m_\psic(\D)
\east
Using the relations $P(\D)P(A)=P(\D\cap A)$, etc., we find
\[\m_\ppp(\D)=\mup(\D\cap A)=(\mup)_\xpp(\D)\]
and so on.

{\bf 5.8.4}

(i) Let $B$ be any Borel set, then
\beast
\m_{\p^i}(B)=\|P(B)\p^i\|^2&=&
(\p,P(B)\p)+(\p^i-\p,P(B)\p^i)+(\p,P(B)(\p^i-\p))\\
&=&\mup(B)+O(\|\p^i-\p\|)\goto\mup(B)
\east

(ii) $\p^i$ are singular, so there exist sets $A_i$ such that
$\m_{\p^i}(A_i^c)=0$ and $m(A_i)=0$. Now $m(\cup A_i)=0$, for all $i$
$\m_{\p^i}((\cup A_j)^c)\leq\m_{\p^i}(A_i^c)=0$ and by (i),
$\mup((\cup A_j)^c)=0$. So $\mup$ is
concentrated on a set of zero Lebesgue measure and, hence, is singular.
If all the $\m_{\p^i}$ are {\em pp} or all are {\em sc} then $\mup$ is
{\em pp} or {\em sc}, respectively, because $\hpp$ and $\hcont$ are closed
subspaces.

{\bf 5.8.5}

(i) $[\hac]P$ is {\em ac}: Let $B$ be a Borel set, $m(B)=0$. Take any
$\p\in\hilb$, then
\[\|[\hac]P(B)\p\|^2=\|P(B)[\hac]\p\|^2=\m_\pac(B)=(\mup)_\xac(B)=0\]
therefore $[\hac]P(B)=0$.

(ii) $[\hsc]P\leq [\hcont]P=\Pcont$, therefore
$[\hsc]P$ is a continuous measure. We have to prove that it is
singular. Choose an orthonormal basis $\{\p_i\}$. Let $A_i$ be sets of zero
Lebesgue measure, $\m_{\p_i^\xsc}$ concentrated on $A_i$. Take $A=\cup A_i$.
$m(A)=0$ and we show that $[\hsc]P$ is concentrated on $A$.
\beast
\|P(A^c)[\hsc]\p_i\|=\|P(A^c)\p_i^\xsc\|
=\|P(\cap_j A_j^c)\p_i^\xsc\|=\|\prod_{j\neq i}P(A_j^c)P(A_i^c)\p_i^\xsc\|=0
\east
The last equality holds because $P(A_i^c)\p_i^\xsc=0$, according to Problem 5.8.1.
It follows that $[\hsc]P(A^c)\p_i=0$ for all $i$, and therefore
$[\hsc]P(A^c)=0$.

{\bf 5.8.6}

If $\m(B)=0$ then $\mup(B)=0$ for all $\p\in\hilb$. Accordingly,
$\|P(B)\p\|=0$ for all $\p\in\hilb$ which means that $P(B)=0$.

{\bf 5.8.7}

Let $B$ be a Borel set, $\m_{\p^1}(B)+\m_{\p^2}(B)=0$. Then
$\|P(B)\p^1\|=\|P(B)\p^2\|=0$ and hence $P(B)\p^1=P(B)\p^2=0$.
This implies
\beast
\m_{\a\p^1+\b\p^2}(B)&=&|\a|^2\|P(B)\p^1\|^2+|\b|^2\|P(B)\p^2\|^2\\
&+&\a^*\b(\p^1,P(B)\p^2)+\a\b^*(\p^2,P(B)\p^1)=0
\east

{\bf 5.8.8}

According to Problem 5.8.6, it suffices to show that $\mup\ll\sum c_i\m_{\p^i}$
for all $\p\in\hilb$. Writing $\p=\sum\a_i\p^i$, one proves
\[\m_{\Sigma\a_i\p^i}\ll\sum c_i\m_{\p^i}\]
exactly as in Problem 5.8.7.

{\bf 5.8.9}

(i) The set is linearly independent: Taking any finite linear combination
with not all coefficients vanishing
and expanding the sum in the canonical basis $\{\d^n\}$,
at least one $\d^n$ appears with nonzero coefficient.

(ii) The set generates the canonical basis $\{\d^n\}_{n=-\infty}^\infty$.
For $n=0$ one obtains $\d^0,\d^1$ and by induction one finds that
$\d^{-n},\d^{-n+1},...,\d^n,\d^{n+1}$ can be expanded with the vectors
$\{H^k\d^0,H^k\d^1\}_{k=0}^n$.

{\bf 5.8.10}

Let $\p^1\in\hilb_1$ and $\p^2\in\hilb_2$ where $\hilb_i$ are subspaces of
$\hilb$.
\beast
\m_{\p^1+\p^2}(B)&=&\|P(B)(\p^1+\p^2)\|^2\\
&=&\m_{\p^1}(B)+\m_{\p^2}(B)+(P(B)\p^1,\p^2)+(\p^2,P(B)\p^1)
\east
and
\[(P(B)\p^1,\p^2)=(P(B)[\hilb_1]\p^1,[\hilb_2]\p^2)=0\]
for every Borel set $B$ if and only if the two subspaces are orthogonal
and one of them, say $\hilb_1$, is $H$-invariant (and, hence, $[\hilb_1]$
commutes with $P(B)$ for all $B$). Then, however,
$\hilb_2\subset\hilb_1^\perp$ which is $H$-invariant and contains $\p^2$.

{\bf 6.2.1}

Choose a sequence $\{n_k\}$ such that $V^k\equiv V(T^{n_k}\o)\goto V(\o')$
pointwise. The corresponding operators $H^k=H_0+V^k$
all have the same spectrum, $\s(H(\o))$,
and converge strongly to $H(\o')$. Equation (\ref{RS}) holds in the form
$\s(H(\o))\supset\s(H(\o'))$. The opposite is also true because $V(\o')$ is
also minimal.

{\bf 6.2.2}

Any point in the spectrum which is not an eigenvalue is in the essential
spectrum. So we have to prove that all the eigenvalues are in the essential
spectrum. Let $E$ be an eigenvalue, $\p$ the corresponding normalized
eigenvector.
Because $V$ is recurrent, there exists an increasing
sequence $\{n_k\}$ such that $V^k=T^{n_k}VT^{-n_k}$
($V^k_n=V_{n+n_k}$) tends to $V$ pointwise. Let $\vf^k=T^{-n_k}\p$. This goes
to zero weakly ($(\vf,\vf^k)\goto0$ for any fixed vector $\vf$), so it is a
Weyl sequence if $\ve^k\equiv (H_0+V-E)\vf^k\goto0$ (in norm!).
\[\|\ve^k\|^2=\|T^{n_k}\ve^k\|^2=\|(H_0+V^k-E)\p\|^2=\|(V^k-V)\p\|^2\]
There exists a sequence of integers $m_k\gotoinf$ and
a sequence $\eta_k\goto0$
such that $|V^k_n-V_n|^2<\eta_k$ for $|n|<m_k$, and
$\sum_{|n|\geq m_k}|\p_n|^2<\eta_k$.
Therefore
\[\|(V^k-V)\p\|^2=(\sum_{|n|<m_k}+\sum_{|n|\geq m_k})(V^k_n-V_n)^2|\p_n|^2
<\eta_k (1+4\sup|V_m|^2)\]
so $\ve^k$ goes to zero, indeed.

{\bf 6.9.1}

According to Section 6.3, the IDS can be obtained as
\[\ids(E)=\lim_{n\gotoinf}\ids_{nL}(E)\]
where $\ids_{nL}(E)$ is $1/nL$ times
the number of eigenvalues less than $E$ of $H_{nL}$,
the restriction of $H$ to an interval of length $nL$.
A convenient choice of boundary condition is
\[\p(k+nL)=e^{i\a}\p(k)\]
The matrix of $H_{nL}$ has $V_1,...,V_{nL}$ in the diagonal, $1$ above and below
it and $e^{-i\a}$ and $e^{i\a}$ in the upper right and lower left corner,
respectively. Finding the eigenvalues of $H_{nL}$ is equivalent to finding
the $E$ values for which the transfer matrix
$T_{1\goto nL}(E)=T_{1\goto L}^n(E)$ has eigenvalues
$e^{\pm i\a}$ or trace $2\cos\a$: this can be seen from
\[\P_{nL}=T_{1\goto nL}\P_0\]
(cf. Section 3.2) and the boundary condition. Choose $\a=\pi/2$, then one
has to look at the zeros of $\tr T_{1\goto L}^n(E)$.
From Figure 1 and Problem 3.2.1 one learns that this function has exactly
$n$ zeros in each band of the spectrum of $H$; therefore the IDS in the gap just
above the $k$th band is $kn/nL=k/L$, independently of $n$, which proves the
assertion.


\section*{References}

\begin{itemize}

\reff[1] J.P.~Allouche:
   This volume.
\eref
\reff[2] J.P.~Allouche, J.~Peyri\`ere:
   \(Sur une formule de r\'ecurrence sur les traces de produits de matrices
   associ\'ees \`a certaines substitutions).\\
   C.R.\ Acad.\ Sci.\ Paris~\>302< s\'erie II, \<1135--1136> (1986).
\eref
\reff[3] Al-Naggar, D.B.~Pearson:
   \(A New Asymptotic Condition for Absolutely Continuous Spectrum
   of the Sturm-Liouville Operator on the Half-Line).\\
   Helv.\ Phys.\ Acta.~\>67<, \<144--166> (1994).
\eref
\reff[4] S.~Aubry, G.~Andr\'e:
   \(Analycity Breaking and Anderson Localisation in Incommensurate Lattices).\\
   Ann.\ Israel Phys. Soc~\>3<, \<133--140> (1980).
\eref
\reff[5] Y.~Avishai, D.~Berend:
   \(Trace Maps for Arbitrary Substitution Sequences).\\
   J.\ Phys.~\>A<: Math.\ Gen.~\>26<, \<2437--2443> (1993).
\eref
\reff[6] J.~Avron, P.H.M.~v.~Mouche, B.~Simon:
   \(On the Measure of the Spectrum for the Almost Mathieu Operator).\\
   Commun.\ Math.\ Phys.~\>132<, \<103--118> (1990).
\eref
\reff[7] J.~Avron, B.~Simon:
   \(Almost Periodic Schr\"odinger Operators I: Limit Periodic\\ Potentials).\\
   Commun.\ Math.\ Phys.~\>82<, \<101--120> (1981).
\eref
\reff[8] J.~Avron, B.~Simon:
   \(Almost Periodic Schr\"odinger Operators II: The Integrated Density of
    States).\\
   Duke Math.\ J.~\>50<, \<369--391> (1983).
\eref
\reff[9]
   F.~Axel, J.P.~Allouche, M.~Kleman, M.~Mendes-France, J.~Peyri\`ere:
   \(Vibrational Modes in a One Dimensional ``Quasi-Alloy'': the Morse Case).\\
   J. de Physique (Paris)~C3 \>47<, \<181--186> (1986).
\eref
\reff[10]
   F.~Axel, J.~Peyri\`ere:
   \(Spectrum and Extended States in a Harmonic Chain with Controlled
     Disorder: Effects of the Thue-Morse Symmetry).\\
   J.~Stat.~Phys.~\>57<, \<1013--1047> (1989).
\eref
\reff[11]
   F.~Axel, J.~Peyri\`ere:
   \(Etats \'etendus dans une cha\^\i ne \`a d\'esordre contr\^ol\'e).\\
   C.R.~Accad.~Sci.~Paris \>306< s\'erie II, \<179--182> (1988).
\eref
\reff[12] J.~Bellissard:
   \(Spectral Properties of Schr\"odinger's Operator with a Thue-Morse
     Po\-tential).\\
   in ``Number Theory and Physics'', ed.~J.M.~Luck, P.~Moussa, M.~Waldschmidt,
   Springer Verlag: Springer Proceedings in Physics~\>47<, \<140--150> (1990).
\eref
\reff[13] J.~Bellissard:
   \(Gap Labelling Theorems for Schr\"odinger Operators).\\
   in ``From Number Theory to Physics'',\\ ed.~M.~Waldschmidt, P.~Moussa,
   J.M.~Luck, C.~Itzykson, chapter~12, \<538--630>
   Springer Verlag, (1992).
\eref
\reff[14] J.~Bellissard:
   \($K$-Theory of $C^*$-Algebras in Solid State Physics).\\
   in ``Statistical Mechanics and Field Theory: Mathematical Aspects'',\\
   ed.~T.C.~Dorlas, N.M.~Hungenholtz, M.~Winnink, Springer Verlag: Lecture Notes
   in Physics~\>257<, \<99--156> (1986).
\eref
\reff[15] J.~Bellissard, A.~Bovier, J.-M.~Ghez:
   \(Spectral Properties of a Tight Binding Hamiltonian with Period Doubling
     Potential).\\
   Commun.\ Math.\ Phys.~\>135<, \<379--399> (1991).
\eref
\reff[16] J.~Bellissard, A.~Bovier, J.-M.~Ghez:
   \(Gap Labelling Theorems for One Dimensional Discrete Schr\"odinger
     Operators).\\
   Rev.~Math.~Phys.~\>4<, \<1--37> (1992).
\eref
\reff[17] J.~Bellissard, B.~Iochum, E.~Scoppola, D.~Testard:
   \(Spectral Properties of One Dimensional Quasi-Crystals).\\
   Commun.~Math.~Phys.~\>125<, \<327--345> (1986).
\eref
\reff[18] J.~Bellissard, B.~Iochum, D.~Testard:
   \(Continuity Properties of the Electronic Spectrum of $1D$ Quasicrystals).\\
   Commun.~Math.~Phys.~\>141<, \<353--380> (1991).
\eref
\reff[19] J.~Bellissard, R.~Lima, D.~Testard:
   \(A Metal-Insulator Transition for the Almost Mathieu Model).\\
   Commun.\ Math.\ Phys.~\>88<, \<207--234> (1983).
\eref
\reff[20] J.~Bellissard, E.~Scoppola:
   \(The Density of States for Almost Periodic Schr\"o\-din\-ger
     Operators and the Frequency Module: A Counter-Example).\\
   Commun.~Math.\ Phys.~\>85<, \<301--308> (1982).
\eref
\reff[21] J.~Bellissard, B.~Simon:
   \(Cantor Spectrum for the Almost Mathieu Equation).\\
   J.~Func.\ Anal.~\>48<, \<408--419> (1982).
\eref
\reff[22] Ju.M.~Berezanskii:
   \(Expansions in Eigenfunctions of Selfadjoint Operators).\\
   A.M.S.~Translations of Mathematical Monographs~\>17<, (1968).
\eref
\reff[23] V.~Berth\'e:
   This volume.
\eref
\reff[24] A.S.~Besicovitch:
   \(Almost Periodic Functions).\\
   Cambridge University Press, (1932).
\eref
\reff[25] Ph.~Bougerol, J.~Lacroix:
   \(Products of Random matrices with Application to Schr\"o\-dinger
     Operators).\\
   Birkh\"auser: Progress in Probability and Statistics, (1985).
\eref
\reff[26] A.~Bovier, J.-M.~Ghez:
   \(Spectral Properties of One-Dimensional Schr\"odinger Operators with
     Potentials Generated by Substitutions).\\
   Commun.\ Math.\ Phys.~\>158<, \<45--66> (1993).
\eref
\reff[27] A.~Bovier, J.-M.~Ghez:
   Erratum to \(Spectral Properties of One-Dimensional Schr\"odinger
   Operators with Potentials Generated by Substitutions),\\
   Commun.\ Math.\ Phys.~\>158<, \<45--66> (1993).\\
   Commun.\ Math.\ Phys.~\>166<, \<431--432> (1994).
\eref
\reff[28] R.~Carmona, A.~Klein, F.~Martinelli:
   \(Anderson Localization for Bernoulli and Other Singular Potentials).\\
   Commun.\ Math.\ Phys.~\>108<, \<41--66> (1987).
\eref
\reff[29] R.~Carmona, J.~Lacroix:
   \(Spectral Theory of Random Schr\"odinger Operators).\\
   Birkh\"auser: Probability and its Applications,  (1990).
\eref
\reff[30] M.~Casdagli:
   \(Symbolic Dynamics for the Renormalization Map of a Quasi\-periodic
   Schr\"odinger Equation).\\
   Commun.~Math.~Phys.~\>107<, \<295--318> (1986).
\eref
\reff[31] M.-D.~Choi, G.A.~Elliot, N.~Yui:$\!\!$
   \(Gauss Polynomials and the Rotation Algebra).
   Invent.\ Math~\>99<, \<225--246> (1990).
\eref
\reff[32] V.~Chulaevsky, F.~Delyon:
   \(Purely Absolutely Continous Spectrum for Almost
   Ma\-thieu Operators).\\
   J.~Stat.\ Phys.~\>55<, \<1279--1284> (1989).
\eref
\reff[33] V.A.~Chulaevsky, Ya.G.~Sinai:
   \(Anderson Localisation for $1$-$D$ Discrete Schr\"o\-din\-ger Operator
   with Two-Frequency Potential).\\
   Commun.\ Math.\ Phys.~\>125<, \<91--112> (1989).
\eref
\reff[34] V.A.~Chulaevsky, Ya.G.~Sinai:
   \(The Exponential Localization and Structure of the Spectrum for $1D$
   Quasi-Periodic Discrete Schr\"odinger Operators).\\
   Rev.\ Math.\ Phys.~\>3<, \<241--284> (1991).
\eref
\reff[35] W.~Craig, B.~Simon:
   \(Subharmonicity of the Lyapunov Index).\\
   Duke Math.\ J.~\>50<, \<551--560> (1983).
\eref
\reff[36] W.~Craig, B.~Simon:
   \(Log H\"older continuity of the Integrated Density of States for
   Stochastic Jacobi Matrices).\\
   Commun.\ Math.\ Phys.~\>90<, \<207--218> (1983).
\eref
\reff[37] H.L.~Cycon, R.G.~Froese, W.~Kirsch, B.~Simon:
   \(Schr\"odinger Operators with Application to Quantum Mechanics and Global
     Geometry).\\
   Springer Verlag, (1987).
\eref
\reff[38] P.~Deift, B.~Simon:
   \(Almost Periodic Schr\"odinger Operators III: The Absolutely Continuous
   Spectrum in One Dimension).\\
   Commun.\ Math.\ Phys.~\>90<, \<389--411> (1983).
\eref
\reff[39] F.M.~Dekking:
   \(The Spectrum of Dynamical Systems Arising from Substitutions of
   Constant Length).\\
   Z.~Wahrscheinlichkeitstheorie verw.\ Gebiete~\>41<, \<221--239> (1978).
\eref
\reff[40] F.M.~Dekking:
   This volume.
\eref
\reff[41] F.~Delyon:
   \(Abscence of  Localization for the Almost Mathieu Equation).\\
   J.~Phys.~\>A<: Math.\ Gen.~\>20<, \<L21--L23> (1987).
\eref
\reff[42] F.~Delyon, D.~Petritis
   \(Absence of Localization in a Class of Schr\"odinger Operators with
   Quasiperiodic Potential).\\
   Commun.~Math.~Phys.~\>103<, \<441--444> (1986).
\eref
\reff[43] F.~Delyon, J.~Peyri\`ere:
   \(Recurrence of the Eigenstates of a Schr\"odinger Operator with Automatic
     Potential).\\
   J.\ Stat.\ Phys.~\>64<, \<363--368> (1991).
\eref
\reff[44] F.~Delyon, B.~Souillard:
   \(The Rotation Number for Finite Difference Operators and its Properties).\\
   Commun.\ Math.\ Phys.~\>89<, \<415--426> (1983).
\eref
\reff[45] M.S.P.~Eastham:
   \(The spectral Theory of Periodic Differential Equations).\\
   Scottish Academic Press, (1973).
\eref
\reff[46] P.~Erd\"os, R.C.~Herndon:
   \(Theories of Electrons in One-Dimensional Disordered Systems).\\
   Adv.\ Phys.~\>31<, \<65--163> (1982).
\eref
\reff[47] H.~Furstenberg, H.~Kesten:
   \(Products of Random Matrices).\\
   Ann.\ Math.\ Stat.~\>31<, \<457--469> (1960).
\eref
\reff[48] S.~Fishman, D.R.~Grempel, R.E.~Prange:
   \(Localization in an Incommensurate Potential: An Exactly Solvable Model).\\
   Phys.\ Rev.\ Lett.~\>49<, \<833--836> (1982).
\eref
\reff[49] S.~Fishman, D.R.~Grempel, R.E.~Prange:
   \(Localization in a $d$-Dimensional Incommensurate Structure).\\
   Phys.\ Rev.~\>B~29<, \<4272--4276> (1984).
\eref
\reff[50] J.~Fr\"ohlich, T.~Spencer, P.~Wittwer:
   \(Localization for a Class of One Dimensional Quasi-Periodic Schr\"odinger
    Operators).\\
   Commun.\ Math.\ Phys.~\>132<, \<5--25> (1990).
\eref
\reff[51] D.J.~Gilbert:
   \(On Subordinacy and Analysis of the Spectrum of Schr\"odinger Operators
   with Two Singular Endpoints).\\
   Proc.\ Royal.\ Soc.\ Edinburgh~\>112A<, \<213--229> (1989).
\eref
\reff[52] D.J.~Gilbert, D.B.~Pearson:
   \(On Subordinacy and Analysis of the Spectrum of One-Dimensional
     Schr\"odinger Operators).\\
   J.~Math.\ Anal.\ and Appl.~\>128<, \<30--56> (1987).
\eref
\reff[53] I.Ya.~Goldsheid:
   \(Asymptotic Properties of the Product of Random Matrices Depending on a
     Parameter).\\
   in \(Multicomponent Random Systems) edited by R.L. Dobrushin\\
   and Ya.G.~Sinai, Marcel Dekker Inc., \<239--283> (1980).
\eref
\reff[54] A.Ya.~Gordon:
   \(On the Point Spectrum of the One Dimensional Schr\"odinger Operator).\\
   Usp.\ Math.\ Nauk.~\>31<, \<257> (1976).
\eref
\reff[55] A.Ya.~Gordon:
   \(Pure Point Spectrum Under $1$-Parameter Perturbations and Instability
    of Anderson Localization).\\
   Commun.\ Math.\ Phys.~\>164<, \<489--505> (1994).
\eref
\reff[56] W.H.~Gottschalk:
   \(Substitution Minimal Sets).\\
   Transactions\ Amer.\ Math.\ Soc.~\>109<, \<467--491> (1963).
\eref
\reff[57] G.H.~Hardy, E.M.~Wright:
   \(An Introduction to the Theory of Numbers).\\
   Fourth Edition, Oxford University Press, (1971).
\eref
\reff[58] D.~Herbert, R.~Jones:
   \(Localized States in Disordered Systems).\\
   J.~Phys.~\>C 4<, \<1145--1161> (1971).
\eref
\reff[59] M.R.~Herman:
   \(Une m\'ethode pour minorer les exposants de Lyapounov et quelques exemples
     montrant le caract\`ere local d'un th\'eor\`eme d'Arnold et de Moser sur
     le tore de dimension $2$).\\
     Comment.\ Math.\ Helvetici~\>58<, \<453--502> (1983).
\eref
\reff[60] H.~Hiramoto, M.~Kohmoto:
   \(New Localization in a Quasiperiodic System).\\
   Phys.\ Rev.\ Lett.~\>62<, \<2714--2717> (1989).
\eref
\reff[61] H.~Hiramoto, M.~Kohmoto:
   \(Electronic Spectral and Wavefunction Properties of One-Dimensional
   Quasiperiodic Systems: A Scaling Approach).\\
   Int.\ J.~of Mod.\ Phys.~\>B~6<, \<281--320> (1992).
\eref
\reff[62] H.~Hiramoto, M.~Kohmoto:
   \(Scaling Analysis of Quasiperiodic Systems: Generalized Harper Model).\\
   Phys.\ Rev.~\>B~40<, \<8225--8234> (1989).
\eref
\reff[63] A.~Hof:
   \(Some Remarks on Discrete Aperiodic Schr\"odinger Operators).\\
   J.~Stat.\ Phys.~\>72<, \<1353--1374> (1993).
\eref
\reff[64] A.~Hof, O.~Knill, B.~Simon:
   \(Singular Continuous Spectrum for Palindromic Schr\"odinger Operators).\\
   Preprint~(1994).
\eref
\reff[65] K.~Iguchi:
   \(Equivalence Between the Nielsen and the Scaling Transformations in
     One-Dimensional Quasiperodic Systems).\\
   J.\ Math.\ Phys.~\>34<, \<3481--3490> (1993).
\eref
\reff[66] B.~Iochum, D.~Testard:
   \(Power Law Growth for the Resistance in the Fibonacci Model).\\
   J.~Stat.~Phys.~\>65<, \<715--723> (1991).
\eref
\reff[67] B.~Iochum, L.~Raymond, D.~Testard:
   \(Resistance of One-Dimensional\\ Quasicrystals).\\
   Physica~\>A 187<, \<353--368> (1992).
\eref
\reff[68] K.~Ishii:
   \(Localization of Eigenstates and Transport Phenomena in the One
     Dimensional Disordered System).\\
   Suppl.\ Prog.\ Theor.\ Phys.~\>53<, \<77--138> (1973).
\eref
\reff[69] S.Ya.~Jitomirskaya:
   \(Anderson Localization for the Almost Mathieu Equation I:
     A Nonperturbative Proof).\\
   Commun.\ Math.\ Phys.~\>165<, \<49--57> (1994).
\eref
\reff[70] S.Ya.~Jitomirskaya:
   \(Anderson Localization for the Almost Mathieu Equation II:
     Point Spectrum for $\lambda>2$).\\
   Preprint (1994).
\eref
\reff[71] S.~Jitomirskaya, B.~Simon:
   \(Operators with Singular Continuous Spectrum:\\
    III:~Almost Periodic Schr\"odinger Operators).\\
   Commun.\ Math.\ Phys.~\>165<, \<201--205> (1994).
\eref
\reff[72] R.~Johnson:
   \(A Review of Recent Works on Almost Periodic Differential and Difference
     Operators).\\
     Acta Appl.\ Math.~\>1<, \<54--78> (1983).
\eref
\reff[73] R.~Johnson, J.~Moser:
   \(The Rotation Number for Almost Periodic Potentials).\\
   Commun.\ Math.\ Phys.~\>84<, \<403--438> (1982).\\
   Erratum: Commun.\ Math.\ Phys.~\>90<, \<317--318> (1983).
\eref
\reff[74] G.~Jona-Lasinio, F.~Martinelli, E.~Scoppola:
   \(Multiple Tunneling in $d$-Dimen\-sion of a Quantum Particle
   in a Hierarchical System).\\
   Ann.\ Inst.\ Henri Poincar\'e~\>42<, \<73--108> (1985).
\eref
\reff[75] I.S.~Kac:
   \(On the Multiplicity of the Spectrum of a Second Order Differential
     Operator).\\
   Soviet Math.~\>3<, \<1035--1039> (1962).
\eref
\reff[76] I.S.~Kac:
   \(On the Multiplicity of the Spectrum of a Second Order Differential
     Operator).\\
   Izv.\ Akad.\ Nauk SSSR Ser. Mat.~\>27<, \<1081--1112> (1963) (in Russian).
\eref
\reff[77] T.~Kato:
   \(Perturbation Theory for Linear Operators).\\
   Springer Verlag: Grund.~der~math.~Wissen.~\>132< 2nd ed., 2nd print,
   (1984).
\eref
\reff[78] S.~Khan, D.B.~Pearson:$\!\!$
   \(Subordinacy and Spectral Theory for Infinite\\ Matrices).\\
   Helv.\ Phys.\ Acta~\>65<, \<505--527> (1992).
\eref
\reff[79] M.~Kohmoto, L.P.~Kadanoff, C.~Tang:
   \(Localization Problem in One Dimension: Mapping and Escape).\\
   Phys.~Rev.~Lett.~\>50<, \<1870--1872> (1983).
\eref
\reff[80] S.~Ostlund, R.~Pandit, D.~Rand, H.J.~Schnellnhuber, E.D.~Siggia:
   \(One-Dimen\-sion Schr\"odinger Equation
   with an Almost Periodic Potential).\\
   Phys.~Rev.~Lett.~\>50<, \<1873--1876> (1983).
\eref
\reff[81] M.~Kohmoto, Y.~Oono:
   \(Cantor Spectrum for an Almost Periodic Schr\"odinger Equation and a
   Dynamical Map).\\
   Phys.~Lett.~\>102 A<, \<145--148> (1984).
\eref
\reff[82] M.~Kohmoto, B.~Sutherland, K.~Iguchi:
   \(Localisation in Optics: Quasiperiodic Media).\\
   Phys.\ Rev.\ Lett.~\>58<, \<2436--2438> (1987).
\eref
\reff[83] M.~Kol\'a\v r, M.K.~Ali:
   \(Trace Maps Associated with General Two-Letter Substitution Rules).\\
   Phys. Rev.~\>A~42<, \<7112--7124> (1990).
\eref
\reff[84] M.~Kol\'a\v r, F.~Nori:
   \(Trace Maps of General Substitutional Sequences).\\
   Phys. Rev.~\>B~42<, \<1062--1065> (1990).
\eref
\reff[85] J.~Koll\'ar, A.~S\"ut\H o:
   \(The Kronig-Penney Model on a Fibonacci Lattice).\\
   Phys.~Lett.~\>117 A<, \<203--209> (1986).
\eref
\reff[86] S.~Kotani:
   \(Lyapunov Indices Determine Absolute Continuous Spectra of Stationary
   One Dimensional Schr\"odinger Operators).\\
   Proc.\ Ky\=oto Stoch.\ Conf.~(1983).
\eref
\reff[87] S.~Kotani:
   \(Lyapunov Exponents and Spectra for One Dimensional Random
   Schr\"odinger Operators).\\
   Proceedings of the A.M.S.\ meeting on `Random Matrices' Brunswick~(1984).
\eref
\reff[88] S.~Kotani:
   \(Jacobi Matrices with Random Potential taking Finitely Many Values).\\
   Rev.\ Math.\ Phys.~\>1<, \<129--133> (1989).
\eref
\reff[89] H.~Kunz:
   This volume.
\eref
\reff[90] H.~Kunz, R.~Livi, A.~S\"ut\H o:
   \(Cantor Spectrum and Singular Continuity for a Hierarchical Hamiltonian).\\
   Commun.\ Math.\ Phys.~\>122<, \<643--679> (1989).
\eref
\reff[91] H.~Kunz, B.~Souillard:
   \(Sur le spectre des op\'erateurs aux diff\'erences finies al\'ea\-toires).\\
   Commun.\ Math. Phys.~\>78<, \<201--246> (1980).
\eref
\reff[92] Y.~Last:
   \(On the Measure of Gaps and Spectra for Discrete $1D$ Schr\"odinger
     Operators).\\
   Commun.\ Math.\ Phys.~\>149<, \<347--360> (1992).
\eref
\reff[93] Y.~Last:
   \(A Relation Between a.c.~Spectrum of Ergodic Jacobi Matrices and the
   Spectra of Periodic Approximants).\\
   Commun.\ Math.\ Phys.~\>151<, \<183--192> (1993).
\eref
\reff[94] Y.~Last:
   \(Zero Measure Spectrum for the Almost Mathieu Operator).\\
   Commun.\ Math.\ Phys.~\>164<, \<421--432> (1994).
\eref
\reff[95] P.~Liardet:
   This volume.
\eref
\reff[96] R.~Livi, A.~Maritan, S.~Ruffo:
   \(The Spectrum of a $1$-$D$ Hierarchical Model).\\
   J.\ Stat.\ Phys.~\>52<, \<595--608> (1988).
\eref
\reff[97] R.~Livi, A.~Politi, S.~Ruffo:
   \(Repeller Structure in a Hierarchical Model: I. Topological Properties).\\
   J.\ Stat.\ Phys.~\>65<, \<53--72> (1991).
\eref
\reff[98] R.~Livi, A.~Politi, S.~Ruffo:
   \(Repeller Structure in a Hierarchical Model: II. Metric Properties).\\
   J.\ Stat.\ Phys.~\>65<, \<73--95> (1991).
\eref
\reff[99] J.M.~Luck:
   \(Cantor Spectra and Scaling of Gap Widths in Deterministic Aperiodic
     Systems).\\
   Phys.~Rev.~\>B 39<, \<5834--5849> (1989).
\eref
\reff[100] V.A.~Mandelshtam, S.Ya.~Zhitomirskaya:
   \($1D$-Quasiperiodic Operators.\\ Latent Symmetries).\\
   Commun.\ Math.\ Phys.~\>139<, \<589--604> (1991).
\eref
\reff[101] F.~Martinelli, E.~Scoppola:
   \(Introduction to the Mathematical Theory of Anderson Localization).\\
   Rivista del nuovo cimento~\>10<, \<1--90> (1987).
\eref
\reff[102] P.~Michel:
   \(Stricte ergodicit\'e d'ensembles minimaux de substitutions).\\
   C.\ R.\ Acad.\ Sci.\ Paris S\'erie~\>A-B 278<, \<811--813> (1974).
\eref
\reff[103] J.~Moser:
   \(An Example of a Schr\"odinger Equation with Almost Periodic Potential
   and Nowhere Dense Spectrum).\\
   Comment.\ Math.\ Helvetici~\>56<, \<198--224> (1981).
\eref
\reff[104] V.I.~Oseledec
   \(A Multiplicative Ergodic Theorem, Ljapunov Characteristic
   Numbers for Dynamical Systems).\\
   Trans.\ Moscow Math.\ Soc.~\>19<, \<197--231> (1968).
\eref
\reff[105] S.~Ostlund, S.~Kim:
   \(Renormalisation of Quasiperiodic Mappings).\\
   Physica Scripta~\>T 9<, \<193--198> (1985).
\eref
\reff[106] L.A.~Pastur:
   \(Spectral Properties of Disordered Systems in the One Body\\
   Approximation).\\
   Commun.\ Math.\ Phys.~\>75<, \<179--196> (1980).
\eref
\reff[107] L.~Pastur, A.~Figotin:
   \(Spectra of Random and Almost-Periodic Operators).\\
   Springer Verlag: Grund.~der~math.~Wissen.~\>297<, (1992).
\eref
\reff[108] D.B.~Pearson:
   \(Quantum Scattering and Spectral Theory).\\
   Academic Press, (1988).
\eref
\reff[109] D.B.~Pearson:
   \(Singular Continuous Measures in Scattering Theory.)\\
    Commun.\ Math.\ Phys.~\>60<, \<13--36> (1978).
\eref
\reff[110] J.~Peyri\`ere:
   \(On the Trace Map for Products of Matrices Associated with Substitutive
     Sequences).\\
   J.\ Stat.\ Phys.~\>62<, \<411--414> (1991).
\eref
\reff[111] J.~Peyri\`ere:
   This volume.
\eref
\reff[112] J.~Peyri\`ere, Z.-Y.~Wen, Z.-X.~Wen:
   \(Polyn\^omes associ\'es aux endomorphismes de groupes libres).\\
   L'Enseignement Math\'ematique~\>39<, \<153--175> (1993).
\eref
\reff[113] M.~Queff\'elec:
   \(Substitution Dynamical Systems. Spectral Analysis).\\
   Springer Verlag: Lecture Notes in Math.~\>1294<, (1987).
\eref
\reff[114] M.~Queff\'elec:
   This volume.
\eref
\reff[115] M.~Reed, B.~Simon:
   \(Methods of Modern Mathematical Physics.\\
     I : Functional Analysis.\\
     IV: Analysis of Operators).\\
   Academic Press, (1980).
\eref
\reff[116] N.~Riedel:
   \(Point Spectrum for the Almost Mathieu Equation).\\
   C.R.\ Math.\ Rep.\ Acad.\ Sci.\ Canada~\>VIII<, \<399--403> (1986).
\eref
\reff[117] N.~Riedel:
   \(Almost Mathieu Operators and Rotation $C^*$-Algebras).\\
   Proc.\ London Math.\ Soc.~\>56<, \<281--302> (1988).
\eref
\reff[118] N.~Riedel:
   \(Absence of Cantor Spectrum for a Class of Schr\"odinger Operators).\\
   Bull.\ of the A.M.S.~\>29<, \<85--87> (1993).
\eref
\reff[119] N.~Riedel:
   \(The Spectrum of a Class of Almost Periodic Operators).\\
   Preprint (1993).
\eref
\reff[120] N.~Riedel:
   \(Regularity of the Spectrum for the Almost Mathieu Operator).\\
   Preprint (1993).
\eref
\reff[121] R.~Riklund, M.~Severin, Y.~Liu:
   \(The Thue-Morse Aperiodic Crystal, a Link Between the Fibonacci
     Quasicrystal and the Periodic Crystal).\\
   Int.~J.~Mod.\ Phys.~\>B~1<, \<121--132> (1987).
\eref
\reff[122] R.\ del~Rio, N.\ Makarov, B.\ Simon:
   \(Operators with Singular Continuous Spec\-trum:
   II.~Rank One Operators).\\
   Commun.\ Math.\ Phys.~\>165<, \<59--67> (1994).
\eref
\reff[123] D.~Ruelle
   \(Ergodic Theory of Differentiable Dynamical Systems).\\
   Publ.\ Math.\ IHES~\>50<, \<275--306> (1979).
\eref
\reff[124] S.~Saks:
   \(Theory of the Integral).\\
   Dover Pub. Inc., (1964).
\eref
\reff[125] T.~Schneider, D.~Wurtz, A.~Politi, M.~Zannetti:
   \(Schr\"odinger Problem for Hierarchical Heterostructures).\\
   Phys.\ Rev.~\>B~36<, \<1789--1792> (1987).
\eref
\reff[126] M.A.~Shubin:
   \(Discrete Magnetic Laplacian).\\
   Commun.\ Math.\ Phys.~\>164<, \<259--275> (1994).
\eref
\reff[127] B.~Simon:
   \(Almost Periodic Schr\"odinger Operators: A Review).\\
   Adv.\ in Appl.\ Math.~\>3<, \<463--490> (1982).
\eref
\reff[128]  B.~Simon:
   \(Schr\"odinger Semigroups).\\
   Bull.\ Amer.\ Math.\ Soc.~\>7<, \<447--526> (1982).
\eref
\reff[129]  B.~Simon:
   \(Kotani Theory for One Dimensional Stochastic Jacobi Matrices).\\
   Commun.\ Math.\ Phys.~\>89<, \<227--234> (1983).
\eref
\reff[130] B.~Simon:
   \(Almost Periodic Schr\"odinger Operators IV: The Maryland Model).\\
   Annals of Phys.~\>159<, \<157--183> (1985).
\eref
\reff[131] Ya.G.~Sinai:
   \(Anderson Localization for One-Dimensional Difference Schr\"odin\-ger
   Operator with Quasiperiodic Potential).\\
   J.~Stat.\ Phys.~\>46<, \<861--909> (1987).
\eref
\reff[132] E.~Sorets, T.~Spencer:
   \(Positive Lyapunov Exponents for Schr\"odinger Operators with Quasi-%
   Periodic Potentials).\\
   Commun.\ Math.\ Phys.~\>142<, \<543--566> (1991).
\eref
\reff[133] A.~Soshnikov:
   \(Difference Almost-Periodic Schr\"odinger Operators: Corollaries of
     Localization).\\
   Commun.\ Math.\ Phys.~\>153<, \<465--477> (1983).
\eref
\reff[134] B.~Sutherland, M.~Kohmoto:
   \(Resistance of a One-Dimensional Quasicrystal: Power Law Growth).\\
   Phys.\ Rev.~\>B~36<, \<5877--5886> (1987).
\eref
\reff[135] A.~S\"ut\H o:
   \(The Spectrum of a Quasiperiodic Schr\"odinger Operator).\\
   Commun.\ Math.\ Phys.~\>111<, \<409--415> (1987).
\eref
\reff[136] A.~S\"ut\H o:
   \(Singular Continuous Spectrum on a Cantor Set of Zero Lebesgue Measure for
   the Fibonacci Hamiltonian).\\
   J.~Stat.~Phys.~\>56<, \<525--531> (1989).
\eref
\reff[137] D.~Thouless:
   \(A relation Between the Density of States and Range of Localization for
     One-Dimensional Random Systems).\\
   J.\ Phys.~\>C~5<, \<77--81> (1972).
\eref
\reff[138] D.~Thouless:
   \(Bandwidths for a Quasiperiodic Tight Binding Model).\\
   Phys.\ Rev.~\>B~28<, \<4272--4276> (1983).
\eref
\reff[139] D.~Thouless:
   \(Scaling for the Discrete Mathieu Equation).\\
   Commun.\ Math.\ Phys.~\>127<, \<187--193> (1990).
\eref
\reff[140] M.~Toda:
   \(Theory of Nonlinear Lattices).\\
   Springer Verlag: Springer Series in Solid-State Sciences~\>20<, (1981).
\eref

\end{itemize}

\end{document}